\documentclass[12pt]{article}
\usepackage{amsfonts, amsmath, amssymb, cite, graphicx, }

\usepackage{color}
\usepackage[all, knot]{xy}
\usepackage{tikz}
\usepackage{subfigure}
\usepackage{braket}

\usepackage[utf8]{inputenc}
\usepackage{ulem}

\usepackage{bm}
\usepackage{epstopdf}
\usepackage[footnotesize]{caption}
\usepackage{amsthm}
\usepackage{amsthm,amsmath,amssymb}
\usepackage{enumitem}
\usepackage{mathrsfs}
\usepackage{makecell}
\usepackage{diagbox}

\usepackage[margin=3cm]{geometry}

\begin{document}

\begin{titlepage}
\vspace{0.5cm}
\begin{center}
{\Large \bf Impacts of wormholes on false vacuum bubble tunneling}

\lineskip .75em
\vskip 2.5cm
{\large Hong Wang$^{a}$, Yuxuan Wu$^{b}$, Ran Li$^{c}$, and Jin Wang$^{d,}$\footnote{Corresponding author, jin.wang.1@stonybrook.edu} }
\vskip 2.5em
 {\normalsize\it $^{a}$State Key Laboratory of Electroanalytical Chemistry, Changchun Institute of Applied Chemistry, Chinese Academy of Sciences, Changchun 130022, China\\$^{b}$Center of Theoretical Physics, College of Physics, Jilin University, Changchun 130012, China\\$^{c}$Department of Physics, Qufu Normal University, Qufu 273165, China\\
 $^{d}$Department of Chemistry and Department of Physics and Astronomy, State University of New York at Stony Brook, NY 11794, USA}
\vskip 3.0em
\end{center}
\begin{abstract}
In this study, we generalize the work of Farhi, Guth and Guven [Nucl. Phys. B {\bf339} (1990) 417] by incorporating a wormhole and investigating the impact of the topological structure of spacetime on false vacuum bubble tunneling. In our model, the spherically symmetric bubble contains two domain walls. The classical dynamics of each domain wall are constrained by two classically forbidden regions. We find that the presence of a wormhole increases the number of instantons, thereby enhancing the tunneling rate. We analytically derive the tunneling rate formula, which reduces to the result obtained by Farhi, Guth, and Guven in the limit where the wormhole vanishes. Our results show that the tunneling rate increases with both the wormhole throat radius and the black hole mass. We demonstrate that the topological structure of spacetime influences domain wall tunneling. Within a limited range of parameters, we find that the tunneling rate increases with surface tension, while it decreases with the cosmological constant. In the case where a  wormhole exists within a bubble and is connected to another universe, we demonstrate that the bubble can enter into another universe through quantum tunneling.
\end{abstract}
\end{titlepage}

\baselineskip=0.7cm

\tableofcontents
\newpage
\section{Introduction}
\label{sec:0}
The false vacuum bubble dynamics have been extensively studied for many years~\cite{EAJ,KMHK,HMKK,K,VVI,SEA,JC,AME,SF,WDJ,SFVF}. For a spherically symmetric bubble, the simplest dynamics is the expansion or contraction, which can be described by the bubble radius. Under the thin-wall approximation, the surface energy density of the domain wall is equal to the surface tension~\cite{SEA}. Neither of them is changed with time~\cite{SEA}. Inside the domain wall is the dS spacetime if the cosmological constant is positive (In this study, we do not consider the case where the cosmological constant is negative). Outside the domain wall is the Schwarzschild spacetime~\cite{EAJ,SEA}. The domain wall can be viewed as the glue that sticks the dS spacetime and Schwarzschild spacetime together. The junction condition provides the bubble dynamical equation, which can be derived from the Einstein equations~\cite{W,SEA}.

The dynamics of the spherically symmetric bubble are similar to those of a one-dimensional particle moving in a potential that resembles a mountain. If the mass of the bubble exceeds the critical value, then the radius of the bubble can increase from  zero to infinity. If the mass is smaller than the critical value, there are two classically allowed regions separated by a classically forbidden region~\cite{EAJ,SEA,WDJ}. Thus, there are two types of classical trajectories. One starts from the white hole singularity, expands to the critical point, and then bounces off  to the black hole singularity. The other contracts from infinity to the critical point and then bounces off back to infinity~\cite{SFVF,WDJ}. Quantum mechanics permits the bubble tunneling from one classically allowed region into another region. This makes the so-called ``free lunch process" possible~\cite{WDJ,EAJ}. In other words, it is possible for the universe to tunnel from a singular point~\cite{WDJ,EAJ,JS,A1,A2}.

There are different methods for studying the quantum tunneling of bubble. In~\cite{SF}, Coleman and De Luccia used Euclidean instanton methods to study the vacuum decay. They showed that different dS false vacuum bubbles can tunnel into each other. The tunneling rate is determined by the Euclidean action of the bounce solution and the background state. In~\cite{EAJ}, Farhi, Guth and Guven (FGG) used Euclidean path integral methods to show that a small bubble can tunnel into a larger bubble and subsequently develop into a new universe.  In~\cite{WDJ}, Fischer, Morgan and Polchinski obtained a similar result by solving the Wheeler-DeWitt equation. Other significant theoretical progresses have been made in relation to bubble tunneling, such as  the tunneling of thermal bubbles~\cite{AD1,AD2,OJ} or charged AdS bubbles~\cite{RJ}, the transition of Minkowski spacetime~\cite{SFVF} and so on,  please refer to~\cite{SFVF,OJ,RJ,dc,vf,jj,31RI,31PR,PG16,OY16,P16,HAT16} and the references therein. We will not list them all here.

The current cosmological observations suggest a homogenous and isotropic universe with a flat space slice ($k=0$)~\cite{SW,SD}. The topology of the universe is assumed to be trivial. However, there is no physical principle that prevents the existence of the genus (or other topological structures)~\cite{MJ}. And the quantum fluctuations of spacetime in the early universe may have produced various topologically non-trivial structures~\cite{SWH,AB}. Some of these structures may still exist today. Thus, it is unnatural to think the topology of this vast universe is trivial. These topological structures of the universe can have observable effects that  may be seen in future experiments~\cite{MJ}. In addition,  with the advancement of the experimental techniques, it is possible to simulate vacuum decay in ultracold atom systems~\cite{JMHAS,KBMMP}. It is also possible to simulate the wormhole in the experiment~\cite{DAJDS,C1}. The topology of a wormhole is often non-trivial. For instance, in three dimensional spacetime, the topology of an Euclidean wormhole is a genus~\cite{kb}. Thus, one expects that it is also possible to experimentally simulate the influence of topology on bubble tunneling in the foreseeable future. Therefore, it is important to theoretically study the influence of topology on bubble tunneling.

Different topological structures may have varying influences on bubble tunneling. The topology of the wormhole is non-trivial~\cite{kb,KK1221}. For simplicity, in this work, we study whether the topology of a wormhole can influence domain wall tunneling.  Specifically,  we generalize the model studied in~\cite{EAJ} to include a wormhole. We derived the classical dynamic equation of the domain walls.  In our model, the bubble has two domain walls. These two domain walls may collide with each other at the throat of the wormhole. The classical motion of each domain wall is constrained by two classically forbidden regions. On each side of the throat, there exists a classically forbidden region. The dynamical potential of the domain wall on different side of the throat is different.  The dynamics of these two domain walls are equivalent to each other. Thus,  it is sufficient to study the dynamics of just one of the domain walls.

Usually, the presence of the genus can increase the number of instantons and then enhance the tunneling rate~\cite{S1,L1}. We show that in our model, the wormhole also increases the number of instantons. In this sense, the role of the wormhole topology in our model is similar to the role of the genus in the model studied in~\cite{S1}. In each classically forbidden region, there exists one type of instanton.  Each pair of instanton-anti-instanton (I-A-I) is an Euclidean bounce~\cite{SC1,SC2}. We calculated the subtracted tunneling action of the instanton. We numerically simulated the relationship between the subtracted tunneling action and various intrinsic properties of the bubble.

Using the Euclidean instanton methods, we have derived the tunneling rate of the bubble. We show that the tunneling rate increases with the throat radius of the wormhole. We also show that the tunneling rate increases with the mass of the black hole. Within a limited range of parameters, the tunneling rate increases with surface tension, but decreases with the cosmological constant.  As the throat radius of the wormhole approaches zero, the tunneling rate approaches the result obtained by FGG. We demonstrate that the topological structure of spacetime can indeed influence the tunneling of the domain wall. Throughout this study, we use units where $G=\hbar=c=1$.


\section{Schwarzschild surgery}
\label{sec:1}
With the advancement of holographic gravity theory~\cite{JM,AK1,AK2,SJ,SS,JV,JD,AS},  wormhole physics has attracted increasing attention~\cite{DAJDS,C1,PD,JX,SEM,ZZ}. One of the methods for constructing a wormhole solution to the Einstein equations is through Schwarzschild surgery~\cite{MV}. We will illustrate the main steps of this surgery using the following simple example.

The Schwarzschild spacetime is given as
\begin{equation}
\label{eq:1.1}
ds^{2}=-A_{s}dt^{2}+A_{s}^{-1}dr^{2}+r^{2}(d\theta^{2}+sin^{2} \theta d\phi^{2}),
\end{equation}
where
\begin{equation}
\label{eq:1.2}
A_{s}=1-\frac{2M}{r}.
\end{equation}
In equation \eqref{eq:1.2}, $M$ is the mass of the black hole. At first, prepare two copies of the manifold \eqref{eq:1.1}. Then, delete the region $r< r_{o}$ (where $r_{o}$ is a positive number) from each manifold. Finally, paste the two manifolds $r\geq r_{o}$ along the boundaries where $r=r_{o}$.  After completing these surgeries, one obtains a spacetime that contains a wormhole, and $r_{o}$ is the radius of the throat of the wormhole~\cite{MV}. These are the main steps of the Schwarzschild surgery. In reference~\cite{MV}, there is a constraint that  $r_{o}>2M$. This condition ensures that  the wormhole is traversable. However, if one does not require the wormhole to be traversable, then there is no need for this constraint.   It is possible to construct other wormhole solutions to the Einstein equations using this surgery~\cite{SWK,JFS,DCD}. The wormhole constructed in reference~\cite{DCD} is non-traversable.

\begin{figure}[tbp]
\centering
\includegraphics[width=9cm]{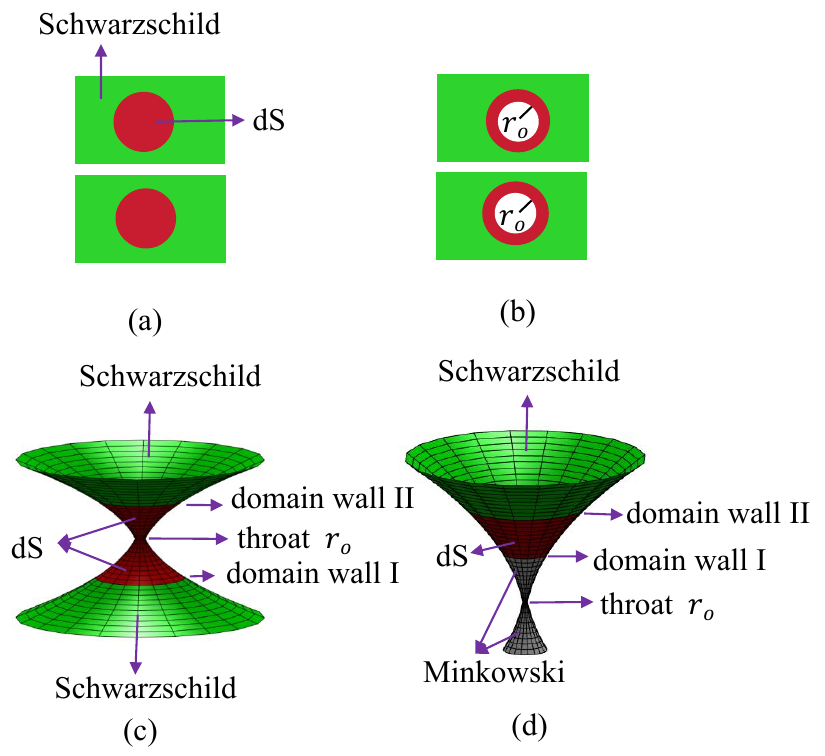}
\caption{\label{fig:1} Taking the Schwarzschild surgery on the spacetime defined by equation \eqref{eq:1.3} creates a wormhole. The green sections depict the Schwarzschild spacetime. The red sections depict the dS spacetime. In figure (b), the white sections with radius $r_{o}$ are deleted. The grey sections in figure (d) represent the Minkowski spacetime.  In figure (c),  there is a domain wall on each side of the throat. In figure (d), the two domain walls are on the same side of the throat. }
\end{figure}

If there is a spherically symmetric false vacuum bubble in spacetime, the solution of the Einstein equations is~\cite{SEA}
\begin{eqnarray}
\label{eq:1.3}
  ds^{2}=\begin{cases}
      -A_{s}dt_{s}^{2}+A_{s}^{-1}dr^{2}+r^{2}(d\theta^{2}+sin^{2} \theta d\phi^{2}),r>r_{1}\\
     -A_{D}dt_{D}^{2}+A_{D}^{-1}dr^{2}+r^{2}(d\theta^{2}+sin^{2} \theta d\phi^{2}),r<r_{1}
   \end{cases},
\end{eqnarray}
where
\begin{equation}
\label{eq:1.4}
A_{D}=1-\chi^{2}r^{2}.
\end{equation}
In equation \eqref{eq:1.4}, $\chi^{2}=8\pi\Lambda/3$ and $\Lambda$ is the cosmological constant. $(t_{s}, r, \theta, \phi)$ and $(t_{D}, r, \theta, \phi)$ are the Schwarzschild coordinates and de Sitter static coordinates, respectively. In equation \eqref{eq:1.3}, we assumed that the coordinate radius of the domain wall is $r_{1}$, which is a dynamical variable of the bubble. This equation shows that outside the domain wall is Schwarzschild spacetime, while inside the domain wall is dS spacetime. In this scenario, the mass of the black hole is equal to the total energy of the bubble. The induced metric $h_{ab}$ ($a, b=0, 1, 2$) on the domain wall is~\cite{RJ}
\begin{equation}
\label{eq:1.5}
ds^{2}=-d\tau^{2}+r_{1}^{2}(d\theta^{2}+sin^{2} \theta d\phi^{2}),
\end{equation}
where, $\tau$ is the proper time measured in the domain wall. According to equation \eqref{eq:1.5}, $h_{00}=h^{00}=-1$, $h_{11}=(h^{11})^{-1}=r_{1}^{2}$ and $h_{22}=(h^{22})^{-1}=r_{1}^{2}sin^{2}\theta$. Other components of the metric $h_{ab}$ (or $h^{ab}$) are zero.

The Schwarzschild surgery can also be applied to the spacetime defined by equation \eqref{eq:1.3}. Firstly, prepare two copies of the manifold defined by equation \eqref{eq:1.3}, as shown in
figure~\ref{fig:1}(a). Next, delete the regions where $r<r_{o}$ for each manifold (figure~\ref{fig:1}(b)). We require that $0<r_{o}<r_{1}$. Finally, paste the two manifolds $r\geq r_{o}$ along the boundaries where $r=r_{o}$. After completing these surgeries, one can obtain a new spacetime. The new spacetime remains spherically symmetric and includes a wormhole, as depicted in figure~\ref{fig:1}(c). The throat of the wormhole has a radial coordinate of $r=r_{o}$,  which is the smallest radial coordinate. The topology of the throat is $S^{2}\times \mathbb{R}$, where $S^{2}$ represents the sphere and $\mathbb{R}$ represents the time direction. After performing an analytic continuation on the time variable ($\tau\rightarrow - i \tau_{E}$), $\mathbb{R}$ becomes $S^{1}$ (circle). Thus, the topology of the Euclidean throat is $S^{2}\times S^{1}$. Noted that the two dimensional torus is $S^{1}\times S^{1}$. Therefore, the Euclidean throat might be viewed as a higher dimensional torus.

There is a domain wall located at $r=r_{1}$ in each side of the throat. Thus, there are two domain walls in the spacetime. Inside the two domain walls is dS spacetime, while outside the two domain walls is the Schwarzschild spacetime.  Thus, we assert that the spherically symmetric bubble possesses two domain walls if there exists a wormhole. For convenience, we define the moment of completing the Schwarzschild surgery as the initial time. Thus in the initial time, there is a domain wall on each side of the throat, as shown in figure~\ref{fig:1}(c). However, with the evolution of the bubble, the two domain walls may appear on the same side of the throat, as depicted in figure~\ref{fig:1}(d). The detailed interpretation of figure~\ref{fig:1}(d) and the dynamics of the domain walls will be presented in the next section.

We point out that in the spacetime defined by equation \eqref{eq:1.3},  a wormhole associated with the horizon at $r=2M$ exists. Thus, in the spacetime illustrated in  figure~\ref{fig:1}(c), in addition to the wormhole introduced via Schwarzschild surgery, there are two other wormholes related to the horizons. However, only the wormhole introduced through Schwarzschild surgery affects the dynamics of the bubble. A more detailed explanation of this point is provided in appendix~\ref{sec:A1}. Hereafter, when we refer to a ``wormhole," we specifically mean the one introduced through Schwarzschild surgery.

The bubble, characterized by two domain walls, is constructed via Schwarzschild surgery. In the later, one can see that this is a simple and useful toy model for studying the influence of topology on domain wall tunneling. We note that this type of bubble can  also be produced via a first order phase transition in the early universe~\cite{KH1221,KM1221}. In~\cite{KH1221}, the authors use a similar bubble to study the production of a baby universe.   Some issues related to this type of bubble remain unresolved, such as the problem of the baryon-entropy ratio of the universe nucleated through this bubble~\cite{KH1221}, and what the creation rate of the baby universe nucleated through this type of bubble is. Therefore, it is important to study this type of bubble. Furthermore, spacetimes with multiple boundaries (for example, the spacetime in figure~\ref{fig:1}(c) has two boundaries ) present intriguing puzzles within holographic duality~\cite{JM1225}. This issue has stimulated extensive research~\cite{JM1225,PEO1225A,PEO1225B,MV1225,DJ1225}.

In Schwarzschild surgery, in order to glue two manifolds together into a spacetime, it is necessary to introduce some additional matter at the throat of the wormhole~\cite{MV,SWK,JFS,MKThorne,FSNL}. For a  traversable wormhole, this additional matter is exotic (violating the weak energy condition or other energy conditions )~\cite{MKThorne}. Perhaps some researchers believe that the weak energy condition can not be violated in principle~\cite{MKThorne}, leading them to consider both the exotic matter and the traversable wormhole constructed via Schwarzschild surgery as unphysical. However, the  Casimir effect in quantum field theory does violate the weak energy condition. This indicates that the weak energy condition (and possibly other energy conditions) is not inviolable~\cite{MV,MKThorne,FSNL,CLXY}. Thus, it would be imprudent to dismiss exotic matter as unphysical~\cite{MV,MKThorne,FSNL,CLXY}.
Moreover, in this work, we do not require the wormhole to be  traversable (please see the next section). For a  non-traversable wormhole, the matter used to glue the manifolds need not be exotic~\cite{DCD}. Therefore, one need not worry about whether the weak energy condition (and other energy conditions) can be violated in this study.

For simplicity, we assume that there is no interaction between the bubble (including the domain walls) and  the matter introduced through Schwarzschild surgery to glue the two manifolds. As a result, this matter does not affect the dynamics (including quantum tunneling) of the bubble. Its sole role is to glue the two manifolds into a single spacetime; any other effects it might have are entirely neglected. We also ignore the dynamics of the throat. Additionally, it is well known that a false vacuum bubble can be modeled using a scalar field with a self-interacting potential~\cite{SF}. In principle, one could introduce an interaction between this matter and the scalar field describing the bubble, which could then affect the bubble’s dynamics and tunneling process. However, such a consideration would significantly complicate the model and lies beyond the scope of this study.

\section{Dynamics of the domain wall}
\label{sec:2}

For clarity, we refer to the two domain walls as \uppercase\expandafter{\romannumeral1} and \uppercase\expandafter{\romannumeral2}. The dynamics of each domain wall are determined by the junction condition, which can be derived from the Einstein equations~\cite{SEA}. For the bubble depicted in figure~\ref{fig:1}(c), the interior of each domain wall is dS spacetime, while the exterior is Schwarzschild spacetime. In this case, the junction condition between the dS and Schwarzschild spacetimes yields the dynamical equation for the domain wall. It can be shown that the equation of motion for the domain wall can be written as~\cite{EAJ,SEA}
\begin{equation}
\label{eq:2.8}
\beta_{D1}-\beta_{s}=\kappa r_{1}.
\end{equation}
Here, $\kappa=4\pi\sigma$, and $\sigma$ denotes the surface tension of the domain wall. The quantity $\beta_{D1}$ is defined as $\beta_{D1}\equiv -A_{D}\dot{t}_{D}$, where $\dot{t}_{D}\equiv d t_{D}/d \tau$. Similarly,  $\beta_{s}\equiv A_{s}\dot{t}_{s}$, where $\dot{t}_{s}\equiv d t_{s}/d \tau$. The derivation of equation \eqref{eq:2.8} is provided in appendix~\ref{sec:A1a}. For the case of figure~\ref{fig:1}(c), the dynamics of both domain walls \uppercase\expandafter{\romannumeral1} and \uppercase\expandafter{\romannumeral2} are governed by equation \eqref{eq:2.8}.
Furthermore, this dynamical equation  can be equivalently rewritten as~\cite{EAJ}
\begin{equation}
\label{eq:2.9}
\dot{r}_{1}^{2}+V_{1}(r_{1},M)=-1,
\end{equation}
where
\begin{equation}
\label{eq:2.10}
V_{1}(r_{1}, M)=-(2M)^{\frac{2}{3}}(\chi^{2}+\kappa^{2})^{\frac{1}{3}}\big\{\frac{1}{\alpha r_{1}}+\frac{(1-\alpha^{3}r_{1}^{3})^{2}}{\gamma^{2}\alpha^{4}r_{1}^{4}}\big\}.
\end{equation}
The parameters $\gamma^{2}$ and $\alpha$ are defined as $\gamma^{2}\equiv 4\kappa^{2}(\chi^{2}+\kappa^{2})^{-1}$ and $\alpha\equiv((\kappa^{2}+\chi^{2})/2M)^{1/3}$, respectively.

\begin{figure}[tbp]
\centering
\includegraphics[width=6cm]{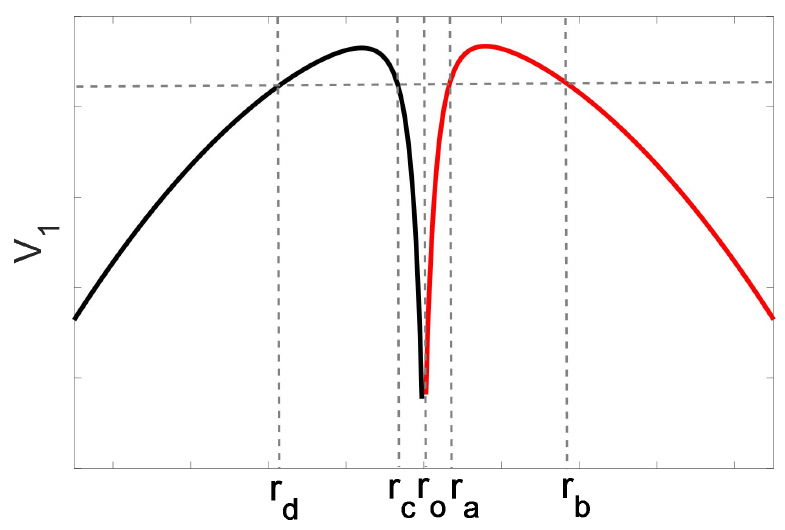}
\caption{\label{fig:2} The diagram of the effective potential $V_{1}$. The curve of different colors represent the potentials of different domain walls. The regions $r_{o}<r_{1}<r_{a}$, $r_{o}<r_{1}<r_{c}$, $r_{1}>r_{b}$ and $r_{1}>r_{d}$ are classically allowed regions. The regions $r_{a}<r_{1}<r_{b}$ and $r_{c}<r_{1}<r_{d}$ are classically forbidden regions. }
\end{figure}

In equation \eqref{eq:2.9}, $V_{1}(r_{1}, M)$ is the effective potential of the domain wall. The variation of $V_{1}$ over $r_{1}$ is shown in figure~\ref{fig:2}. One can show that when~\cite{SEA}
\begin{eqnarray}\begin{split}
\label{eq:2.11}
r_{1}&=\alpha^{-1}2^{-1/3}\big\{[8+(1-\gamma^{2}/2)^{2}]^{1/2}-(1-\gamma^{2}/2)\big\}^{1/3}\\&\equiv r_{m},
\end{split}
\end{eqnarray}
$dV_{1}/dr_{1}=0$. Thus, $V_{1}(r_{m}, M)$ is the maximum value of the potential $V_{1}(r_{1},M)$.
The critical mass $M_{cr}$ is defined as $V_{1}(r_{m}, M_{cr})=-1$. Substituting equation \eqref{eq:2.10} and \eqref{eq:2.11} into the definition of $M_{cr}$, one can show that~\cite{SEA}
\begin{equation}
\label{eq:2.12}
M_{cr}=\frac{4}{3}\pi\chi^{-3}\Lambda\cdot\frac{\gamma^{3}\alpha^{6}r_{m}^{6}(1-\gamma^{2}/4)^{1/2}}{3\sqrt{3}(\alpha^{6}r_{m}^{6}-1)^{3/2}}.
\end{equation}
When  $M>M_{cr}$, there is no classically forbidden region. When $M<M_{cr}$, there are two points which satisfy $\dot{r}_{1}=0$ on one side of the throat. We denote these two points as $r_{a}$ and $r_{b}$. Without loss of generality, we set $r_{a}<r_{b}$. In this study, we always assume that $r_{o}<r_{a}$. The points $r_{a}$ and $r_{b}$ are critical points that separate the classically allowed region from the classically forbidden region. The regions $r_{o}<r_{1}<r_{a}$ and $r_{1}>r_{b}$ are classically allowed regions. The region $r_{a}<r_{1}<r_{b}$ is the classically forbidden region. On the other side of the throat, another domain wall is also under the same potential.  There are also two points satisfy $\dot{r}_{1}=0$. We denote these two points as $r_{c}$ and $r_{d}$ ($r_{c}<r_{d}$).

\begin{figure}[tbp]
\centering
\includegraphics[width=8.5cm]{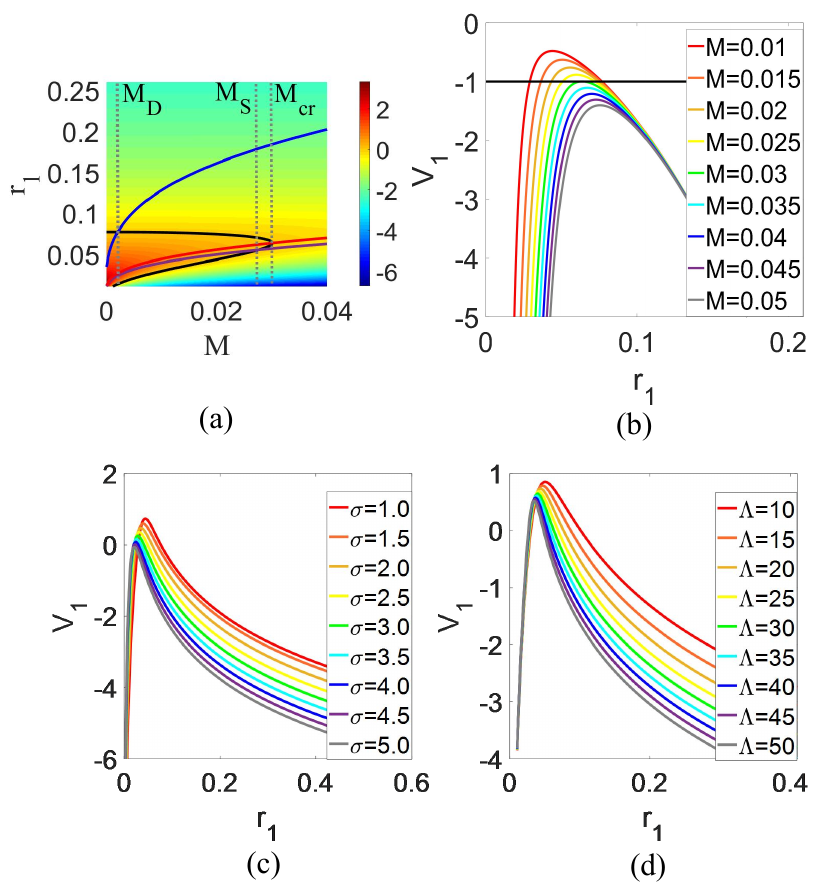}
\caption{\label{fig:3} Diagram of the potential function $V_{1}$. In figure (a), the horizontal axis represents the mass parameter $M$, while the vertical axis represents the radial coordinate $r_{1}$. Various colors indicate different values of $-\mathrm{ln}(-V_{1})$. The black curve is defined as $V_{1}(r_{1},M)=-1$. The red curve is defined as $\partial_{r_{1}}V_{1}=0$. The purple curve is defined as $r_{1}=r_{s}$. The blue curve is defined as $r_{1}=r_{D}$. In figures (b), (c) and (d), the horizontal and vertical axes represent the radial coordinate $r_{1}$ and the potential $V_{1}$, respectively. In figure (b), the parameters are taken as: $\sigma=1$, $\Lambda=20$. In figure (c), the parameters are taken as: $M=0.01$, $\Lambda=20$. In figure (d), the parameters are taken as: $\sigma=1$, $M=0.01$.}
\end{figure}

Figure~\ref{fig:3} illustrates how the potential $V_{1}$  changes under the influence of  certain parameters. In figure~\ref{fig:3}(a), the region enclosed  by the black curve represents the classically forbidden region, while the area outside corresponds to the classically allowed region. In figure~\ref{fig:3}(b), different curves represent various values of the mass $M$. Regions where $V_{1}>-1$ and $V_{1}<-1$ correspond to classically forbidden and allowed regions, respectively. Figures~\ref{fig:3} (c) and ~\ref{fig:3}(d) show that different values of the surface tension $\sigma$ or the cosmological constant $\Lambda$ correspond to similar shapes of the potential $V_{1}$. This suggests that different $\sigma$ (or $\Lambda$) correspond to similar dynamical behavior of the domain wall.

The zero points of the parameters $\beta_{s}$ and $\beta_{D1}$ are crucial for deriving the subtracted tunneling action.  One can prove that the parameters $\beta_{s}$ and $\beta_{D1}$ can be expressed as~\cite{SEA}
\begin{equation}
\label{eq:2.14}
\beta_{s}=(2M)^{\frac{1}{3}}(\chi^{2}+\kappa^{2})^{\frac{2}{3}}\cdot\frac{1-\alpha^{3}r_{1}^{3}}{2\kappa\alpha^{2}r_{1}^{2}},
\end{equation}
and
\begin{equation}
\label{eq:2.15}
\beta_{D1}=(2M)^{\frac{1}{3}}(\chi^{2}+\kappa^{2})^{\frac{2}{3}}\cdot\frac{1-(1-\gamma^{2}/2)\alpha^{3}r_{1}^{3}}{2\kappa\gamma^{2}r_{1}^{2}},
\end{equation}
respectively. Equation \eqref{eq:2.14} shows that when $r_{1}=\alpha^{-1}\equiv r_{s}$,  $\beta_{s}=0$. If $\gamma^{2}\geq 2$, $\beta_{D1}$ is always positive. If $\gamma^{2}<2$, $\beta_{D1}$ can be  either positive or negative. In this study, we always constrain that $0<\gamma^{2}<2$. According to equation \eqref{eq:2.15}, one can easily show that when $r_{1}=\alpha^{-1}(1-\gamma^{2}/2)^{-1/3}\equiv r_{D}$,  $\beta_{D1}=0$.

Based on the definitions of $r_{s}$ and $r_{D}$, it is useful to introduce the parameters $M_{s}$ and $M_{D}$ via the definitions  $V_{1}(r_{s}, M_{s})=-1$ and $V_{1}(r_{D}, M_{D})=-1$, respectively. Then one can  show that~\cite{SEA}
\begin{equation}
\label{eq:2.16}
M_{s}=\frac{4}{3}\pi\chi^{-3}\Lambda(1-\gamma^{2}/4)^{1/2}
\end{equation}
and
\begin{equation}
\label{eq:2.17}
M_{D}=\frac{4}{3}\pi\chi^{-3}\Lambda\cdot\frac{1-\gamma^{2}/2}{1-\gamma^{2}/4}.
\end{equation}
In the case where $0<\gamma^{2}<2$, one can prove that $M_{D}<M_{s}<M_{cr}$.

We assume that at the initial time (the moment of completing the schwarzschild surgery), domain wall \uppercase\expandafter{\romannumeral1} and \uppercase\expandafter{\romannumeral2} are located at $r_{o1}$ and $r_{o2}$, respectively. The values of $r_{o1}$ and $r_{o2}$ are assumed to be the same, yet they are distributed in different sides of the throat. The initial state of the bubble is assumed to be  $r_{o}<r_{o1}<r_{a}$ and  $r_{o}<r_{o2}<r_{c}$, with $\dot{r}_{o1}>0$ and $\dot{r}_{o2}>0$. Thus, in the initial state, domain wall \uppercase\expandafter{\romannumeral1} is expanding and located in the classically allowed region $r_{o}<r_{1}<r_{a}$.

We consider the case where the mass parameter $M$ is smaller than the critical mass $M_{cr}$. According to equation \eqref{eq:2.9}, \eqref{eq:2.10} and figure~\ref{fig:2}, domain wall \uppercase\expandafter{\romannumeral1} will expand to reach the critical point $r_{a}$. Once the radius of domain wall \uppercase\expandafter{\romannumeral1} reaches $r_{a}$, its radial velocity will be zero. The domain wall \uppercase\expandafter{\romannumeral1} will then contract until its radius reaches  $r_{o}$. Equation \eqref{eq:2.9} shows that the radial velocity of the domain wall  is not zero at this point. Classically, the two domain walls can  collide at the position of the throat. The collision of the domain walls is usually complicated and may result in the radiation of gravitational waves, the creation of black holes, and other phenomena~\cite{ZCW,ST,RTM,JJL1,JJL2,JJL3,RM,MK,TJLE}. In this study, we consider the simplest case where the collision is elastic. This means that the collision does not change the energy of the domain wall, but it does reverse its radial velocity. After the collision, the bubble will expand once again.

However, semiclassically,  it is  possible that domain wall \uppercase\expandafter{\romannumeral1} can tunnel into the classically allowed region $r_{1}>r_{b}$  through the Euclidean instanton trajectories in the classically forbidden region $r_{a}<r_{1}<r_{b}$. The instanton trajectories in the region $r_{a}<r_{1}<r_{b}$ can be described by the Euclidean version of the dynamical equation \eqref{eq:2.9}. That is $\dot{r}_{1,E}^{2}-V_{1}(r_{1},M)=1$, where $\dot{r}_{1,E}=dr_{1}/d\tau_{E}$ and $\tau_{E}=i\tau$. Domain wall \uppercase\expandafter{\romannumeral2} may  tunnel into the classically allowed region $r_{1}>r_{d}$ when domain wall \uppercase\expandafter{\romannumeral1} reaches the position of the throat. Thus, semiclassically, it is possible that the two domain walls will not collide at the throat. Therefore, it is possible for domain wall \uppercase\expandafter{\romannumeral1} to pass through the throat and emerge on the other side.

When domain wall \uppercase\expandafter{\romannumeral1} passes through the throat, the two domain walls will emerge on the same side of the throat, as shown in figure~\ref{fig:1}(d). The radius of domain wall \uppercase\expandafter{\romannumeral1} is smaller than the radius of domain wall \uppercase\expandafter{\romannumeral2}. (The bubble in figure~\ref{fig:1}(d) resembles a hollow sphere, and  the gravitational field is zero in the empty region  within the sphere.) Inside domain wall \uppercase\expandafter{\romannumeral1} and on the other side of the throat, the cosmological constant is zero.  Thus, in these regions, the spacetime is Minkowski spacetime, and the metric is $ds^{2}=-dt_{M}^{2}+dx^{2}+dy^{2}+dz^{2}$. Here, $(t_{M}, x, y, z)$ are Cartesian coordinates.  More strictly speaking, the spacetime in these regions is not Minkowski spacetime if we consider the influence of the matter introduced during the Schwarzschild surgery to glue the two manifolds together into a single spacetime. However, for simplicity, we neglect the influence of this matter and approximate the spacetime in these regions as Minkowski spacetime. With this simplification, the tunneling action corresponding to figure~\ref{fig:1}(d) can be analytically derived. Please refer to section~\ref{sec:3.2} for further details. The region between domain walls  \uppercase\expandafter{\romannumeral1} and  \uppercase\expandafter{\romannumeral2} is the dS spacetime with the metric $ds^{2}=-A_{D}dt_{D}^{2}+A_{D}^{-1}dr^{2}+r^{2}(d\theta^{2}+sin^{2} \theta d\phi^{2})$.  Thus, in this case, equation \eqref{eq:2.8} can not be used to describe the dynamics of domain wall \uppercase\expandafter{\romannumeral1}.

In the case of figure~\ref{fig:1}(d), the junction condition between Minkowski spacetime and  dS spacetime determines the dynamics of domain wall \uppercase\expandafter{\romannumeral1}.   By applying the same method used to derive equation \eqref{eq:2.8}, one can obtain the dynamical equation of domain wall \uppercase\expandafter{\romannumeral1} as follows:
\begin{equation}
\label{eq:2.18}
\beta_{M}-\beta_{D2}=\kappa r_{1}.
\end{equation}
Here,
\begin{equation}
\label{eq:2.19}
\beta_{M}\equiv \dot{t}_{M}=\frac{1}{2}(\kappa+\frac{\chi}{\kappa})r_{1}
\end{equation}
and
\begin{equation}
\label{eq:2.20}
\beta_{D2}\equiv -A_{D}\dot{t}_{D}=\frac{1}{2}(-\kappa+\frac{\chi}{\kappa})r_{1}.
\end{equation}
Noted that both $\beta_{D1}$ and $\beta_{D2}$ are formally defined as $-A_{D}\dot{t}_{D}$, yet they are different, as shown by equations \eqref{eq:2.15} and \eqref{eq:2.20}. This is caused by the fact that in different cases, the variation of $r_{1}$ with respected to the proper time variable $\tau$ is different.

We have demonstrated that domain wall \uppercase\expandafter{\romannumeral1} may pass through the throat.  For the same reason, it is also possible that domain wall  \uppercase\expandafter{\romannumeral2} passes through the throat, resulting in both domain walls being located on the same side of it. In such a case, the radius of domain wall \uppercase\expandafter{\romannumeral2} is smaller than that of domain wall \uppercase\expandafter{\romannumeral1}. The interior of domain wall \uppercase\expandafter{\romannumeral2} is Minkowski spacetime, the region between the two domain walls is dS spacetime, and the region outside domain wall \uppercase\expandafter{\romannumeral1} is Schwarzschild spacetime. Thus, in this situation, the dynamics of domain walls \uppercase\expandafter{\romannumeral1} and \uppercase\expandafter{\romannumeral2} are  governed by equations \eqref{eq:2.8} and \eqref{eq:2.18}, respectively.

Overall, when the two domain walls lie on opposite sides of the throat, their dynamics are both governed by equation \eqref{eq:2.8}. When they lie on the same side, as in figure~\ref{fig:1}(d),  their radii differ, and the domain wall with the smaller radius follows equation \eqref{eq:2.18}, while the one with the larger radius follows equation \eqref{eq:2.8}. In other words, the dynamics of any individual domain wall are characterized by equations \eqref{eq:2.8} and \eqref{eq:2.18}. In addition, we have assumed that, at the initial time,  both domain walls have the same radius and thus the same radial velocity. Therefore, it is sufficient to analyze the dynamics of a single domain wall, such as domain wall  \uppercase\expandafter{\romannumeral1}.

The dynamical equation \eqref{eq:2.18} can also be written as
\begin{equation}
\label{eq:2.21}
\dot{r}_{1}^{2}+V_{2}(r_{1})=-1,
\end{equation}
where
\begin{equation}
\label{eq:2.22}
V_{2}(r_{1})=-\frac{(\kappa^{2}+\chi^{2})^{2}}{4\kappa^{2}}r_{1}^{2}\equiv-(\frac{r_{1}}{r_{e}})^{2}.
\end{equation}
In equation \eqref{eq:2.21}, $V_{2}(r_{1})$ is the effective potential of the domain wall \uppercase\expandafter{\romannumeral1}. In equation \eqref{eq:2.22}, we have defined the parameter $r_{e}\equiv2\kappa/(\kappa^{2}+\chi^{2})$.

According to equations \eqref{eq:2.21} and \eqref{eq:2.22}, it is straightforward to show that $r_{1}=r_{e}$ is a critical point. At this point, $\dot{r}_{1}=0$. In the case where $r_{o}<r_{e}$,  the region $r_{o}<r_{1}<r_{e}$ is the classically forbidden region, and the region $r_{1}>r_{e}$ is the classically allowed region. In contrast, when $r_{o}\geq r_{e}$, the potential $V_{2}(r_{1})$ can not provide a classically forbidden region. Thus, in the case where $r_{o}\geq r_{e}$, even without considering quantum effects, the domain wall can go across the throat, followed by expansion as described in equation \eqref{eq:2.21}. In other words, $r_{o}\geq r_{e}$ is the condition that the wormhole is traversable for the domain wall which has dynamics  determined by equation \eqref{eq:2.21}. On the other hand, it is well known that the condition for a wormhole to be traversable for a local object is the absence of a horizon~\cite{MKThorne}. Obviously, these two conditions are different.  Thus, a wormhole that is traversable for a local object does not imply that it is also traversable for the domain wall. In this work, we are interested in the quantum tunneling of the domain wall. Therefore, we focus on the case where $r_{o}<r_{e}$. In this scenario, although classical dynamics prevent the domain wall from crossing the throat,  quantum tunneling allows the domain wall to go across the throat with a certain probability. This is why, in the Schwarzschild surgery, we do not require the wormhole to be traversable for a local object.

\begin{figure}[tbp]
\centering
\includegraphics[width=6cm]{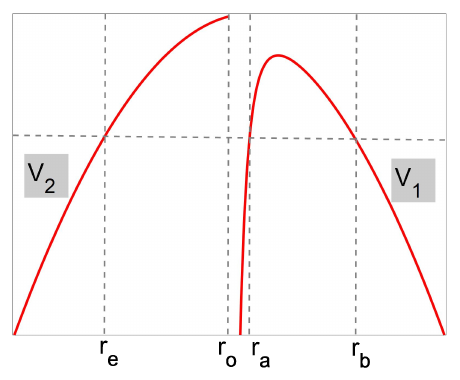}
\caption{\label{fig:6}Diagram of the potential function $V_{1}\bigcup V_{2}$. $r_{o}$ is the radius of the throat. $r_{a}$, $r_{b}$ and $r_{e}$ are critical points of the bubble. The regions $r_{o}<r_{1}<r_{a}$, $r_{1}>r_{b}$ and $r_{1}>r_{e}$ are classically allowed regions. The regions $r_{a}<r_{1}<r_{b}$ and $r_{o}<r_{1}<r_{e}$ are classically forbidden regions.}
\end{figure}

To sum up, the bubble has two domain walls. Each of them has an effective potential $V_{1}$ on one side of the throat and another potential $V_{2}$ on the other side, as shown in figure~\ref{fig:6}. When the mass parameter $M<M_{cr}$, there are three classically allowed regions: $r_{o}<r_{1}<r_{a}$, $r_{1}>r_{b}$ and $r>r_{e}$. The classically forbidden regions are $r_{a}<r_{1}<r_{b}$ and $r_{o}<r_{1}<r_{e}$.  These characteristics are different from the case where there is no wormhole. In the latter case, the bubble only has one domain wall, which only has the effective potential $V_{1}(r_{1})$. And the only classically forbidden region is $r_{a}<r_{1}<r_{b}$. We point out that in each classically forbidden region, there exists one type of instanton. Thus, we assert that the presence of the wormhole has increased the number of instantons.

\section{The subtracted tunneling action}
\label{sec:3}
\subsection{The action $S_{ab}$}
\label{sec:3.1}

In this section, we calculate the subtracted tunneling action, which is necessary for deriving the formula for the tunneling rate. Conventionally, when a one-dimensional particle tunnels in a classically forbidden region, the subtracted tunneling action (Euclidean) is defined as~\cite{EAJ}
\begin{equation}
\label{eq:3.1}
I_{E}\equiv \int_{\tau_{i}}^{\tau_{f}}(L_{E}(x,\dot{x})-L_{E}(x,\dot{x}=0))d\tau_{E}.
\end{equation}
Here, $x$ represents the dynamical variable of the particle. $\tau_{E}$ represents the Euclidean time, while $\tau_{i}$ and $\tau_{f}$ represent the moments when the particle enters and exits the classically forbidden region, respectively. $L_{E}(x,\dot{x})$ is the Euclidean Lagrangian of the particle. $L_{E}(x,\dot{x}=0)$ is the Euclidean Lagrangian at the critical point.

The subtracted tunneling action  $I_{E}$ determines the characteristics of tunneling. Under the WKB approximation, the solution of the Schr$\mathrm{\ddot{o}}$dinger equation is  $e^{\pm I_{E}}$. The sign ambiguity is generally present in quantum mechanics. To describe the tunneling process, one usually chooses the exponential suppressed solution~\cite{EAJ}. In other words, if $I_{E}$ is positive, the tunneling amplitude is $e^{- I_{E}}$, while if $I_{E}$ is negative, the tunneling amplitude is $e^{I_{E}}$. Thus, the tunneling amplitude can be expressed as $e^{-|I_{E}|}$ in any case.  In this study, we follow this convention.

Both equations \eqref{eq:2.9} and \eqref{eq:2.21} show that the dynamics of the domain wall are similar to those of a one-dimensional particle. Thus, we can use a similar definition to equation \eqref{eq:3.1} to calculate the subtracted tunneling action of the domain wall. Specifically, the subtracted tunneling actions of the domain wall in the classically forbidden regions $r_{a}<r_{1}<r_{b}$  and $r_{o}<r_{1}<r_{e}$ are defined as
\begin{equation}
\label{eq:3.2}
S_{ab}\equiv \int_{\tau_{a}}^{\tau_{b}}(L_{Eab}(r_{1},\dot{r}_{1})-L_{Eab}(r_{1},\dot{r}_{1}=0))d\tau_{E}
\end{equation}
and
\begin{equation}
\label{eq:3.3}
S_{oe}\equiv \int_{\tau_{o}}^{\tau_{e}}(L_{Eoe}(r_{1},\dot{r}_{1})-L_{Eoe}(r_{1},\dot{r}_{1}=0))d\tau_{E},
\end{equation}
respectively. Here, $\tau_{a}$ ($\tau_{o}$) and $\tau_{b}$ ($\tau_{e}$) represent the moment when the domain wall enters and exits the classically forbidden region $r_{a}<r_{1}<r_{b}$ ($r_{o}<r_{1}<r_{e}$), respectively. $L_{Eab}(r_{1},\dot{r}_{1})$ ($L_{Eoe}(r_{1},\dot{r}_{1})$) represents the effective Euclidean Lagrangian of the domain wall in the classically forbidden region $r_{a}<r_{1}<r_{b}$ ($r_{o}<r_{1}<r_{e}$). $L_{Eab}(r_{1},\dot{r}_{1}=0)$ (or $L_{Eoe}(r_{1},\dot{r}_{1}=0)$) represents the effective Euclidean Lagrangian at the critical point.

The subtracted tunneling action is defined as the difference between the Euclidean action of the nucleated spacetime configuration and that of the background configuration~\cite{SFVF}. In the Schwarzschild surgery, we introduced additional matter to glue the manifolds together. For simplicity, we neglect the dynamics of both this matter and the wormhole. This implies that, in both the nucleated spacetime configuration and the background configuration, this matter remains static at the throat. As a result, it contributes equally to the Euclidean action in both configurations and therefore does not affect the subtracted tunneling action. Consequently, this matter does not influence the tunneling of the bubble.

\begin{figure}[tbp]
\centering
\includegraphics[width=8cm]{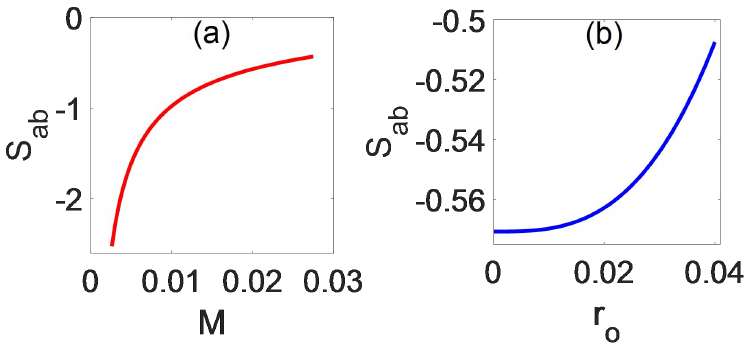}
\caption{\label{fig:7} Variations in the subtracted tunneling action $S_{ab}$ over the mass $M$ and the radius $r_{o}$. The vertical axes of figures (a) and (b) represent the tunneling action $S_{ab}$.  In figure (a), the horizontal axis represents the mass parameter of the Schwarzschild spacetime. The parameters are taken as: $\sigma=1$, $r_{o}=0.01$ and $\Lambda=10$. In figure (b), the horizontal axis represents the radius of the throat. The parameters are taken as: $\sigma=1$, $M=0.02$ and $\Lambda=10$. }
\end{figure}

At first, we calculate the action $S_{ab}$. According to equation \eqref{eq:3.2}, it is convenient to rewrite $S_{ab}$ as
\begin{equation}
\label{eq:3.4}
S_{ab}=I^{tot}_{ab,E}-I^{sub}_{ab,E},
\end{equation}
where $I^{tot}_{ab,E}\equiv \int_{\tau_{a}}^{\tau_{b}}L_{Eab}(r_{1},\dot{r}_{1})d\tau_{E}$ and $I^{sub}_{ab,E}\equiv\int_{\tau_{a}}^{\tau_{b}}L_{Eab}(r_{1},\dot{r}_{1}=0)d\tau_{E}$. The physical interpretation of $I^{tot}_{ab,E}$ is that it represents the Euclidean action of the bubble as it sweeps through the classically forbidden region $r_{a}<r_{1}<r_{b}$. And $I^{sub}_{ab,E}$ represents the subtracted term.

For the gravitational system, the total action is composed of the action of spacetime and the action of matter. FGG has shown that the Euclidean action $I^{tot}_{ab,E}$  of the bubble is composed of four parts~\cite{EAJ}:
\begin{equation}
\label{eq:3.5}
I^{tot}_{ab,E}=I^{wall}_{ab,mat,E}+I^{wall}_{ab,gra,E}+I^{bulk}_{ab,mat,E}+I^{bulk}_{ab,gra,E}.
\end{equation}
Here, $I^{wall}_{ab,mat,E}$ represents the Euclidean action of the domain wall. $I^{wall}_{ab,gra,E}$ represents the Euclidean Einstein-Hilbert action of the infinitesimal neighborhood of the domain wall. $I^{bulk}_{ab,mat,E}$ represents the Euclidean action of the false vacuum inside the domain wall. $I^{bulk}_{ab,gra,E}$ represents the Euclidean Einstein-Hilbert action inside the domain wall. All these actions correspond to the instanton trajectory $r_{a}\rightarrow r_{b}$.  Note that equation \eqref{eq:3.5} is valid irrespect to whether there exist a wormhole or not.

Specifically, the four actions on the right hand side of equation \eqref{eq:3.5} are defined as~\cite{EAJ}:
\begin{equation}
\label{eq:3.6}
I^{wall}_{ab,mat,E}\equiv 4\pi\sigma \int_{\tau_{a}}^{\tau_{b}}r_{1}^{2}d\tau_{E},
\end{equation}

\begin{equation}
\label{eq:3.7}
I^{wall}_{ab,gra,E}\equiv \frac{-1}{16\pi}\lim_{\epsilon\rightarrow0}\int_{-\epsilon}^{+\epsilon}d\eta\int_{\tau_{a}}^{\tau_{b}}d\tau_{E}\int d\theta d\phi r^{2}sin \theta\mathscr{R},
\end{equation}

\begin{equation}
\label{eq:3.8}
I^{bulk}_{ab,mat,E}\equiv-\Lambda\int_{t_{DEa}}^{t_{DEb}}dt_{DE}\int_{r_{o}}^{r_{1}}dr\int d\theta d\phi r^{2}sin \theta
\end{equation}
and
\begin{equation}
\label{eq:3.9}
I^{bulk}_{ab,gra,E}\equiv \frac{-1}{16\pi}\int_{t_{DEa}}^{t_{DEb}}dt_{DE}\int_{r_{o}}^{r_{1}}dr\int d\theta d\phi r^{2}sin \theta\mathscr{R}.
\end{equation}
Here, $\mathscr{R}$ is the Ricci scalar. $t_{DE}$ is the Euclidean de Sitter static coordinate time variable, i.e.,  $t_{DE}=it_{D}$. $t_{DEa}$ and $t_{DEb}$ represent the moment when the domain wall enters and exits the classically forbidden region $r_{a}<r_{1}<r_{b}$, respectively. In equation \eqref{eq:3.7}, $\eta$ represents the proper distance away from the domain wall.   Equations \eqref{eq:3.6} and \eqref{eq:3.8} are the Euclidean actions of the matter. Equations \eqref{eq:3.7} and \eqref{eq:3.9} are the Euclidean Einstein-Hilbert actions.

By substituting equations \eqref{eq:3.5}-\eqref{eq:3.9} into equation \eqref{eq:3.4},  the  action $S_{ab}$ can be obtained as
\begin{eqnarray}\begin{split}
\label{eq:3.19}
S_{ab}=&-2\pi\sigma \int_{\tau_{a}}^{\tau_{b}}r_{1}^{2}d\tau_{E}-\frac{4}{3}\pi\Lambda\int_{\tau_{a}}^{\tau_{b}}d\tau_{E}\frac{\beta_{D1}}{A_{D}}(r_{1}^{3}-r_{o}^{3})\\&-\frac{4}{3}\pi^{2}\Lambda\chi^{-1}(\chi^{-3}-r_{o}^{3})\Theta(M_{D}-M)
\\&-\frac{1}{2}M\big\{\int_{\tau_{a}}^{\tau_{b}}\frac{\beta_{s}}{A_{s}}d\tau_{E}-4\pi M\Theta(M_{s}-M)\big\}\\&-\frac{4}{3}\pi\Lambda r_{o}^{3}\beta_{D1}^{-1}\Big|_{\dot{r}_{1}=0}\int_{\tau_{a}}^{\tau_{b}}d\tau_{E}.
\end{split}
\end{eqnarray}
Here, $\Theta(x)$ denotes the Heaviside step function. A detailed derivation of equation \eqref{eq:3.19} is provided in appendix~\ref{sec:A2}.  It is easy to show that when $r_{o}\rightarrow 0$, the action $S_{ab}$ in equation \eqref{eq:3.19} becomes
\begin{eqnarray}\begin{split}
\label{eq:3.19ma}
S_{ab}\Big|_{r_{o}\rightarrow 0}=&-2\pi\sigma \int_{\tau_{a}}^{\tau_{b}}r_{1}^{2}d\tau_{E}-\frac{4}{3}\pi\Lambda\int_{\tau_{a}}^{\tau_{b}}d\tau_{E}\frac{\beta_{D1}}{A_{D}}r_{1}^{3}\\&-\frac{4}{3}\pi^{2}\Lambda\chi^{-4}\Theta(M_{D}-M)
\\&-\frac{1}{2}M\big\{\int_{\tau_{a}}^{\tau_{b}}\frac{\beta_{s}}{A_{s}}d\tau_{E}-4\pi M\Theta(M_{s}-M)\big\}.
\end{split}
\end{eqnarray}
Equation \eqref{eq:3.19ma} is consistent with equation (5.30) in reference~\cite{EAJ}. Without the Heaviside function related terms, the action $S_{ab}$ can not be a continuous function of the mass parameter $M$~\cite{EAJ}.

It is difficult to analytically perform the integration over the time variable $\tau_{E}$ in equation \eqref{eq:3.19}. Thus, we numerically calculate certain characteristics of the action $S_{ab}$.  Figures~\ref{fig:7}(a) and ~\ref{fig:7}(b) show that the action $S_{ab}$  monotonically increases with respect to the parameters $M$ and $r_{o}$.
Noted that $S_{ab}$ is negative  within  the parameters interval being considered. Thus the tunneling amplitude contributed by the instanton trajectory $r_{a}\rightarrow r_{b}$ is $e^{-|S_{ab}|}$. Therefore, figures~\ref{fig:7}(a) and ~\ref{fig:7}(b) show that the increases in the mass parameter of the black hole or the radius of the throat enhance the tunneling amplitude, which is associated with the instanton trajectory $r_{a}\rightarrow r_{b}$.

\subsection{The action $S_{oe}$}
\label{sec:3.2}

For the subtracted tunneling action $S_{oe}$, which corresponds to the instanton trajectory $r_{o}\rightarrow r_{e}$, it is also composed of the Euclidean action of the bubble on one side of the throat (figure~\ref{fig:1}(d)) and the subtracted term. The Euclidean action of the bubble consists of four parts, similar to the case of equation \eqref{eq:3.5}. Thus, the action $S_{oe}$ can be written as
\begin{equation}
\label{eq:3.20}
S_{oe}=I^{wall}_{oe,mat,E}+I^{wall}_{oe,gra,E}+I^{bulk}_{oe,mat,E}+I^{bulk}_{oe,gra,E}-I^{sub}_{oe,E}.
\end{equation}
Here, $I^{sub}_{oe,E}$ is the subtracted term. It is defined as $I^{sub}_{oe,E}\equiv\int_{\tau_{o}}^{\tau_{e}}L_{Eoe}(r_{1},\dot{r}_{1}=0)d\tau_{E}$. The other actions (such as the surface tension related term of domain wall \uppercase\expandafter{\romannumeral2}) of the bubble do not influence the tunneling of domain wall \uppercase\expandafter{\romannumeral1}. Thus, we have neglected them. Keeping in mind that we always focus on the dynamics of domain wall \uppercase\expandafter{\romannumeral1}.

The actions $I^{wall}_{oe,mat,E}$ and $I^{wall}_{oe,gra,E}$ related to the domain wall under consideration (domain wall \uppercase\expandafter{\romannumeral1}). The actions $I^{bulk}_{oe,mat,E}$ and $I^{bulk}_{oe,gra,E}$ are related to the bulk of the bubble. These four actions  are defined as
\begin{equation}
\label{eq:3.21}
I^{wall}_{oe,mat,E}\equiv 4\pi\sigma \int_{\tau_{o}}^{\tau_{e}}r_{1}^{2}d\tau_{E},
\end{equation}

\begin{equation}
\label{eq:3.22}
I^{wall}_{oe,gra,E}\equiv \frac{-1}{16\pi}\lim_{\epsilon\rightarrow0}\int_{-\epsilon}^{+\epsilon}d\eta\int_{\tau_{o}}^{\tau_{e}}d\tau_{E}\int d\theta d\phi r^{2}sin \theta\mathscr{R},
\end{equation}

\begin{equation}
\label{eq:3.23}
I^{bulk}_{oe,mat,E}\equiv\Lambda\int_{t_{DEo}}^{t_{DEe}}dt_{DE}\int_{r_{o}}^{r_{1}}dr\int d\theta d\phi r^{2}sin \theta
\end{equation}
and
\begin{equation}
\label{eq:3.24}
I^{bulk}_{oe,gra,E}\equiv \frac{1}{16\pi}\int_{t_{DEo}}^{t_{DEe}}dt_{DE}\int_{r_{o}}^{r_{1}}dr\int d\theta d\phi r^{2}sin \theta\mathscr{R}.
\end{equation}
Here, $t_{DEo}$ and $t_{DEe}$ represent the moment when the domain wall enters and exits the classically forbidden region $r_{o}<r_{1}<r_{e}$, respectively.

The definitions \eqref{eq:3.21}-\eqref{eq:3.24} are similar to the definitions \eqref{eq:3.6}-\eqref{eq:3.9}. However, there are certain different aspects. Firstly, the actions \eqref{eq:3.21}-\eqref{eq:3.24} correspond to the instanton trajectory $r_{o}\rightarrow r_{e}$, while the actions \eqref{eq:3.6}-\eqref{eq:3.9} correspond to the instanton trajectory $r_{a}\rightarrow r_{b}$. Additionally, there is a minus sign difference between the definition \eqref{eq:3.23} (\eqref{eq:3.24}) and \eqref{eq:3.8} (\eqref{eq:3.9}).  This difference is caused by the fact that in the case of definitions \eqref{eq:3.8} and \eqref{eq:3.9}, the false vacuum region is located inside the domain wall \uppercase\expandafter{\romannumeral1}. However,  in the case of definitions \eqref{eq:3.23} and \eqref{eq:3.24}, the false vacuum region is located outside the domain wall \uppercase\expandafter{\romannumeral1}. In equation \eqref{eq:3.8}, the factor $-\Lambda\int_{r_{o}}^{r_{1}}dr\int d\theta d\phi r^{2}sin \theta$ implies that with the expansion of the domain wall \uppercase\expandafter{\romannumeral1}, the false vacuum energy increases. This is consistent with the fact that in the case of equation \eqref{eq:3.8}, inside the domain wall \uppercase\expandafter{\romannumeral1} (figure~\ref{fig:1}(c)) is the false vacuum region. With the expansion of the domain wall \uppercase\expandafter{\romannumeral1}, the false vacuum region increases. However, in the case of equation \eqref{eq:3.23}, outside the domain wall \uppercase\expandafter{\romannumeral1} is the false vacuum region (figure~\ref{fig:1}(d)). As the domain wall \uppercase\expandafter{\romannumeral1} expands, the false vacuum region decreases. In other words, the expansion of the domain wall \uppercase\expandafter{\romannumeral1} decreases the false vacuum energy. Thus, the factor  $-\Lambda\int_{r_{o}}^{r_{1}}dr\int d\theta d\phi r^{2}sin \theta$ in the definition of \eqref{eq:3.8} is replaced by $\Lambda\int_{r_{o}}^{r_{1}}dr\int d\theta d\phi r^{2}sin \theta$ in the definition of \eqref{eq:3.23}. Therefore, the action \eqref{eq:3.23} is different from the action \eqref{eq:3.8} by a minus sign. Similar explanations can be used to clarify the minus sign difference between \eqref{eq:3.24} and \eqref{eq:3.9}.

Substituting equations \eqref{eq:3.21}-\eqref{eq:3.24} into equation \eqref{eq:3.20}, one can show that the action $S_{oe}$ is given by
\begin{eqnarray}\begin{split}
\label{eq:3.27}
S_{oe}=&-\pi\sigma r_{e}^{3}\big\{r_{o}r_{e}^{-1}(1-\frac{r_{o}^{2}}{r_{e}^{2}})^{1/2}+arccos(r_{o}r_{e}^{-1})\big\} -\frac{2}{3}\pi\Lambda r_{o}^{3}\chi^{-1}(\frac{\chi}{\kappa}-\kappa)(r_{e}^{-2}-\chi^{2})^{-1/2}\\&\times arctan\big(\frac{\chi(1-r_{o}^{2}r_{e}^{-2})^{1/2}}{(r_{e}^{-2}-\chi^{2})^{1/2}}\big)
+\frac{2}{3}\pi\Lambda(\frac{\chi}{\kappa}-\kappa)\chi^{-4}\Big\{\frac{1}{2}r_{e}^{3}\big\{-(\chi^{2}-2r_{e}^{-2})arcsin(r_{e}^{-1}r_{e})\\&-\chi^{2}r_{e}^{-1}r_{o}(1-r_{e}^{-2}r_{o}^{2})^{1/2}
+(\chi^{2}-2r_{e}^{-2})arcsin(r_{e}^{-1}r_{o})\big\}-2r_{e}arcsin(r_{e}^{-1}r_{e})\\&+2r_{e}arcsin(r_{e}^{-1}r_{o})
-\frac{r_{e}}{2(1-\chi^{2}r_{e}^{2})^{1/2}}\big\{arctan\big(\frac{2r_{o}(r_{e}^{2}-r_{o}^{2})^{1/2}(1-\chi^{2}r_{e}^{2})^{1/2}}{\chi^{2}r_{o}^{2}r_{e}^{2}+r_{e}^{2}-2r_{o}^{2}}\big)-\pi\big\}\Big\}
\\&+\big\{\frac{2}{3}\pi\Lambda r_{o}^{3}r_{e}^{2}(\frac{\chi}{\kappa}-\kappa)(1-\chi^{2}r_{e}^{2})^{-1}-8\pi\sigma r_{e}^{3}\big\}\cdot\big\{arcsin(r_{e}r_{e}^{-1})-arcsin(r_{o}r_{e}^{-1})\big\}.
\end{split}
\end{eqnarray}
A detailed derivation of equation \eqref{eq:3.27} is provided in appendix~\ref{sec:B}.

\section{Tunneling amplitude and tunneling rate}
\label{sec:4}

\subsection{Tunneling amplitude}
\label{sec:4.1}

We assume that at the initial time, domain wall \uppercase\expandafter{\romannumeral1} is located in the classically allowed region $r_{o}<r_{1}<r_{a}$ with the radius $r_{o1}$ and the initial radial velocity $\dot{r}_{o1}>0$. Quantum mechanically, it is possible for domain wall \uppercase\expandafter{\romannumeral1} to pass  through the classically forbidden region $r_{a}<r_{1}<r_{b}$ ($r_{o}<r_{1}<r_{e}$) and then enter the classically allowed region $r_{1}>r_{b}$ ($r_{1}>r_{e}$). In this section, we will study the tunneling rate of the process in which domain wall \uppercase\expandafter{\romannumeral1} tunnels from the region $r_{o}<r_{1}<r_{a}$ into other classically allowed regions.

We denote the initial time and the final time as $-T/2$ and $T/2$, respectively. We use $r_{x}$ to represent any point in the region  $r_{o}<r_{1}<r_{a}$, then $r_{o}<r_{x}<r_{a}$. We use $K(r_{x},\frac{T}{2}; r_{o1}, -\frac{T}{2})$ to represent the amplitude of domain wall \uppercase\expandafter{\romannumeral1} from the state ($r_{o1}$, $-\frac{T}{2}$) evolving to the state ($r_{x}$, $\frac{T}{2}$). All trajectories in the state space which start from the state ($r_{o1}$, $-\frac{T}{2}$) and end in the state ($r_{x}$, $\frac{T}{2}$) contribute to the amplitude $K(r_{x},\frac{T}{2}; r_{o1}, -\frac{T}{2})$. Therefore, we need to sum the contributions of all these trajectories.

We consider this case where the time interval ($-T/2$, $T/2$) is large enough so that  for any number of collisions may have occurred during the evolution process. We can divide all these trajectories into different classes according to the number of collisions. Thus, the amplitude $K(r_{x},\frac{T}{2}; r_{o1}, -\frac{T}{2})$ can be expressed as
\begin{equation}
\label{eq:4.1}
K(r_{x},\frac{T}{2}; r_{o1}, -\frac{T}{2})=\sum_{m=0}^{\infty}K^{(m)}(r_{x},\frac{T}{2}; r_{o1}, -\frac{T}{2}),
\end{equation}
where,  $K^{(m)}(r_{x},\frac{T}{2}; r_{o1}, -\frac{T}{2})$ represents the amplitude contributed by the classes of trajectories in which the two domain walls collide ``$m$'' times during the evolution process, $m=0,1,2,...$

\subsubsection{The amplitude without collision}
\label{sec:4.1.1}

\begin{figure}[tbp]
\centering
\includegraphics[width=8cm]{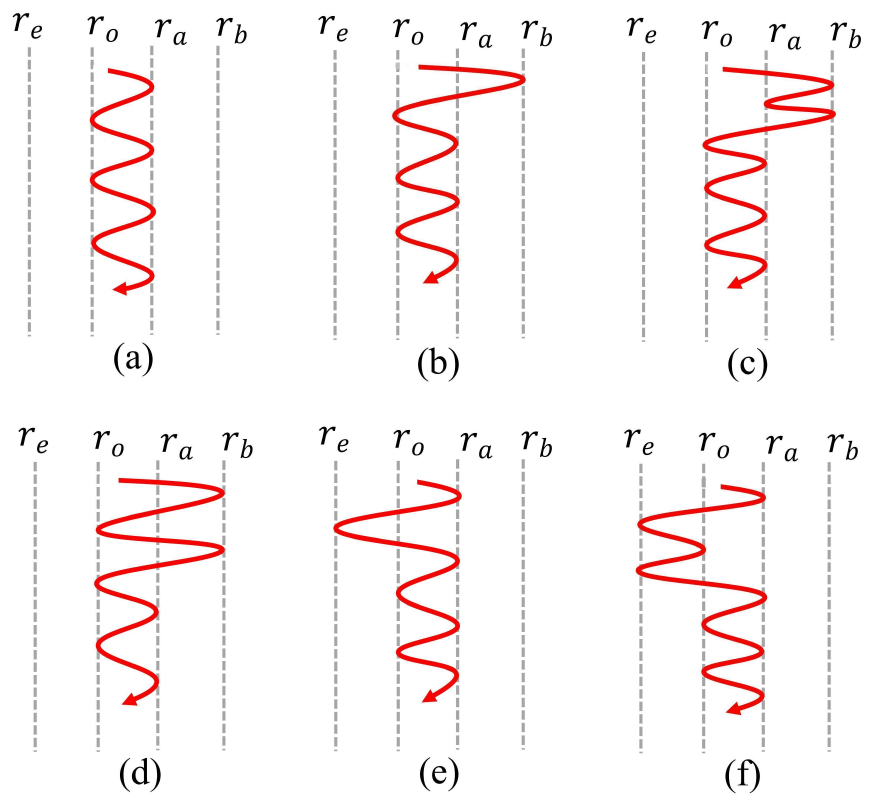}
\caption{\label{fig:9} Diagram of some of the trajectories that contribute to the amplitude of the domain wall being in the classically allowed region $r_{o} < r_{1} < r_{a}$. $r_{o}$ represents the radius of the throat. $r_{a}$, $r_{b}$ and $r_{e}$ are critical points of the bubble. The vertical gray dashed lines indicate isocontours at radii $r_{e}$, $r_{o}$, $r_{a}$, and $r_{b}$, respectively. The red curve in (a) is the classically allowed trajectory, while the red curves in (b)-(f) indicate that various I-A-I pairs are inserted into the classically allowed trajectory}
\end{figure}

Even in the simplest case where the two domain walls do not collide at all, there are  infinite number of trajectories contributing to $K^{(0)}(r_{x},\frac{T}{2}; r_{o1}, -\frac{T}{2})$. We need to sum all these contributions. Some of these trajectories are shown in figure~\ref{fig:9}. In figure~\ref{fig:9}(a), the red curve represents the classically allowed trajectory
\begin{equation}
\label{eq:4.2}
r_{o1}\rightarrow r_{a}\rightarrow r_{o} \rightarrow r_{a}\rightarrow r_{o}\rightarrow\cdot\cdot\cdot\rightarrow r_{a}\rightarrow r_{x}.
\end{equation}
This trajectory represents the oscillation of the domain wall \uppercase\expandafter{\romannumeral1} within the classically allowed region $r_{o}<r_{1}<r_{a}$.
The bounce at the point $r_{o}$ is not caused by the collision of the two domain walls. It is caused by the potential $V_{2}$. And the bounce at the point $r_{a}$ is caused by the potential $V_{1}$. Equation \eqref{eq:4.2} represents that the last time of the bounce has occurred at the point $r_{a}$. In fact, it is also possible that the bounce of the last time has occurred at the point $r_{o}$. However,  as we have assumed that the time interval ($-T/2$, $T/2$) is large. Thus, the number of bounce is large. Consequently, the difference between the trajectory $r_{o}\rightarrow r_{x}$ and $r_{o}\rightarrow r_{a}\rightarrow r_{x}$ can be neglected when compared to the trajectory \eqref{eq:4.2}. In this sense, the classically allowed trajectory can always be represented by equation \eqref{eq:4.2}. Figures~\ref{fig:9}(b) and ~\ref{fig:9}(e) represent that a pair of I-A-I $r_{a}\rightarrow r_{b}\rightarrow r_{a}$ and $r_{o}\rightarrow r_{e}\rightarrow r_{o}$ are inserted into the trajectory \eqref{eq:4.2}, respectively. Figures~\ref{fig:9}(c) and ~\ref{fig:9}(d) represent two pairs of  I-A-I $r_{a}\rightarrow r_{b}\rightarrow r_{a}$ are inserted into \eqref{eq:4.2} using different methods.  Figure~\ref{fig:9}(f) represents that two pairs of I-A-I $r_{o}\rightarrow r_{e}\rightarrow r_{o}$ are inserted into \eqref{eq:4.2}.

The amplitude of the trajectory \eqref{eq:4.2} is
\begin{equation}
\label{eq:4.3}
\triangle_{o1a}e^{-S_{o1a}}\triangle_{oa}^{2(N_{1}-1)}e^{-2(N_{1}-1)S_{oa}}\triangle_{ax}e^{-S_{ax}}.
\end{equation}
Here, $\triangle_{o1a}$ ($\triangle_{oa}^{2}$ or $\triangle_{ax}$) represents the fluctuation around the trajectory $r_{o1}\rightarrow r_{a}$ ($r_{o}\rightarrow r_{a}\rightarrow r_{o}$ or $r_{a}\rightarrow r_{x}$). $S_{o1a}$ ($S_{oa}$ or $S_{ax}$) represents the Euclidean action of the trajectory $r_{o1}\rightarrow r_{a}$ ($r_{o}\rightarrow r_{a}$ or $r_{a}\rightarrow r_{x}$). The factor ($N_{1}-1$) represents the number of times of the trajectory $r_{a}\rightarrow r_{o}\rightarrow r_{a}$ is repeated in trajectory \eqref{eq:4.2}. Equation \eqref{eq:4.3} is expressed in the Euclidean proper time coordinate. Thus, the actions $S_{o1a}$, $S_{oa}$  and $S_{ax}$ all are pure imaginary numbers (i.e., the real part of these actions are zero).

Figure~\ref{fig:9}(b) represents that we inserted an I-A-I pairs ($r_{a}\rightarrow r_{b}\rightarrow r_{a}$) into the trajectory \eqref{eq:4.2} when the domain wall \uppercase\expandafter{\romannumeral1} first arrives at the point $r_{a}$. Noted that the Euclidean action of the instanton and anti-instanton is equivalent to each other~\cite{SC1,SC2,UHPP,RHB}. Thus, the amplitude related to the I-A-I pairs is
\begin{equation}
\label{eq:4.4}
\triangle_{ab}^{2}e^{-2|S_{ab}|}.
\end{equation}
Here, $\triangle_{ab}^{2}$  represents the fluctuation  around the I-A-I pairs. And $2S_{ab}$ represents the Euclidean action of a pair of I-A-I.

Figure~\ref{fig:9}(c) shows that two pairs of I-A-I are inserted into the trajectory \eqref{eq:4.2} when the domain wall \uppercase\expandafter{\romannumeral1} first arrives at the point $r_{a}$. More generally, we can insert any pair of I-A-I into the trajectory \eqref{eq:4.2} when the domain wall first arrives at the point $r_{a}$. The amplitude of all these trajectories is
\begin{eqnarray}\begin{split}
\label{eq:4.5}
&\triangle_{o1a}e^{-S_{o1a}}\triangle_{oa}^{2(N_{1}-1)}e^{-2(N_{1}-1)S_{oa}}\\&\times\triangle_{ax}e^{-S_{ax}}\sum_{m=1}^{\infty}\triangle_{ab}^{2m}e^{-2m|S_{ab}|}.
\end{split}
\end{eqnarray}
In equation \eqref{eq:4.5}, ``$m$'' represents the number of I-A-I pairs, which is inserted into the trajectory \eqref{eq:4.2} when the domain wall \uppercase\expandafter{\romannumeral1} first arrives at the point $r_{a}$.

In trajectory \eqref{eq:4.2}, everytime the domain wall \uppercase\expandafter{\romannumeral1} arrives at the point $r_{a}$, one can insert any pair of I-A-I. In other words, one can insert the factor $\sum_{m=1}^{\infty}\triangle_{ab}^{2m}e^{-2m|S_{ab}|}$ at every point $r_{a}$ of trajectory \eqref{eq:4.2}.
Similarly, one can insert the factor
\begin{equation}
\label{eq:4.6}
\sum_{m=1}^{\infty}\triangle_{oe}^{2m}e^{-2m|S_{oe}|}
\end{equation}
at every point $r_{o}$ of the trajectory \eqref{eq:4.2}. In equation \eqref{eq:4.6}, $\triangle_{oe}^{2}$ represents the fluctuation around a pair of I-A-I  $r_{o}\rightarrow r_{e}\rightarrow r_{o}$. And $2S_{oe}$ represents the Euclidean action of a pair of I-A-I  $r_{o}\rightarrow r_{e}\rightarrow r_{o}$. Noted that the number of the point $r_{a}$ in the trajectory \eqref{eq:4.2} is $N_{1}$. The number of the point $r_{o}$ in the trajectory \eqref{eq:4.2} is $N_{1}-1\approx N_{1}$ (we assumed $T$ is large, thus $N_{1}\gg1$).
Summing all of these  insertion modes (also including the amplitude of the trajectory \eqref{eq:4.2}), One can obtain that the amplitude $K^{(0)}(r_{x},\frac{T}{2}; r_{o1}, -\frac{T}{2})$ as
\begin{eqnarray}\begin{split}
\label{eq:4.7}
K^{(0)}(r_{x},\frac{T}{2}; r_{o1}, -\frac{T}{2})=&\triangle_{o1a}e^{-S_{o1a}}\triangle_{oa}^{2(N_{1}-1)}e^{-2(N_{1}-1)S_{oa}}\\&\times\triangle_{ax}e^{-S_{ax}}\sum_{n=0}^{N_{1}}C_{N_{1}}^{n}
(\sum_{m=1}^{\infty}\triangle_{ab}^{2m}\\&\times e^{-2m|S_{ab}|})^{n}\sum_{n_{1}=0}^{N_{1}}C_{N_{1}}^{n_{1}}\\&\times(\sum_{m_{1}=1}^{\infty}\triangle_{oe}^{2m_{1}}e^{-2m_{1}|S_{oe}|})^{n_{1}}.
\end{split}
\end{eqnarray}
Here, $C_{N}^{n}\equiv N!/[n!(N-n)!]$. In equation \eqref{eq:4.7}, the term of $n=n_{1}=0$ represents the amplitude \eqref{eq:4.3}.

We point out that in deriving equation \eqref{eq:4.7}, we neglected the time spent by the I-A-I pairs $r_{a}\rightarrow r_{b}\rightarrow r_{a}$ or $r_{o}\rightarrow r_{e}\rightarrow r_{o}$. For instance, the trajectories depicted in figure~\ref{fig:9} all correspond to the same initial time of $-T/2$ and final time of $T/2$. In other words, we have approximately treated the instanton or anti-instanton as  a point  on the time axis. This approximation is often employed when studying different tunneling problems using Euclidean instanton methods~\cite{SC2,UHPP,RHB}. In our model, the time of the single I-A-I trajectory $r_{a}\rightarrow r_{b}\rightarrow r_{a}$ or $r_{o}\rightarrow r_{e}\rightarrow r_{o}$ can be neglected when compared to the time $T$. Thus, both the instanton and anti-instanton can be regarded as a single point on the time axis. For the multiple I-A-I trajectory $r_{a}\rightarrow r_{b}\rightarrow r_{a}\rightarrow\cdot\cdot\cdot\rightarrow r_{a}$ ($r_{o}\rightarrow r_{e}\rightarrow r_{o}\rightarrow\cdot\cdot\cdot\rightarrow r_{o}$), one may consider that the time spent by this trajectory can not be neglected when the trajectory is sufficiently long. However, this type of trajectory corresponds to a factor $e^{-2m|S_{ab}|}$ ($e^{-2m|S_{oe}|}$) with $m\gg1$. Consequently, the influence of this type of trajectory on the amplitude $K^{(0)}(r_{x},\frac{T}{2}; r_{o1}, -\frac{T}{2})$ is negligible, regardless of whether we neglect the time spent by the I-A-I pairs or not. Therefore, in our model, we can always neglect the time spent by the instanton or anti-instanton.

For convenience, we define
\begin{equation}
\label{eq:4.8}
Q\equiv\sum_{m=1}^{\infty}\triangle_{ab}^{2m}e^{-2m|S_{ab}|}=\frac{\triangle_{ab}^{2}e^{-2|S_{ab}|}}{1-\triangle_{ab}^{2}e^{-2|S_{ab}|}}
\end{equation}
and
\begin{equation}
\label{eq:4.9}
Y\equiv\sum_{m=1}^{\infty}\triangle_{oe}^{2m}e^{-2m|S_{oe}|}=\frac{\triangle_{oe}^{2}e^{-2|S_{oe}|}}{1-\triangle_{oe}^{2}e^{-2|S_{oe}|}}.
\end{equation}
Substituting equations \eqref{eq:4.8} and \eqref{eq:4.9} into \eqref{eq:4.7}, the amplitude $K^{(0)}(r_{x},\frac{T}{2}; r_{o1}, -\frac{T}{2})$ can be simplified as
\begin{eqnarray}\begin{split}
\label{eq:4.10}
K^{(0)}(r_{x},\frac{T}{2}; r_{o1}, -\frac{T}{2})=&\triangle_{o1a}e^{-S_{o1a}}\triangle_{oa}^{2(N_{1}-1)}\\&\times e^{-2(N_{1}-1)S_{oa}}\triangle_{ax}e^{-S_{ax}}\\&\times(1+Q)^{N_{1}}(1+Y)^{N_{1}}.
\end{split}
\end{eqnarray}
Comparing equation \eqref{eq:4.3} and \eqref{eq:4.10}, one can see that if we insert the factor $(1+Q)^{N_{1}}(1+Y)^{N_{1}}$ into the amplitude \eqref{eq:4.3}, then we can obtain the amplitude $K^{(0)}(r_{x},\frac{T}{2}; r_{o1}, -\frac{T}{2})$. Each point $r_{a}$ and $r_{o}$ on the trajectory \eqref{eq:4.2} corresponds to a factor $(1+Q)$ and $(1+Y)$, respectively.
This indicates that at every point $r_{a}$ ($r_{o}$)  on the trajectory \eqref{eq:4.2}, one can insert any pair of I-A-I $r_{a}\rightarrow r_{b}\rightarrow r_{a}$ ($r_{o}\rightarrow r_{e}\rightarrow r_{o}$).

\subsubsection{The impact of collision}
\label{sec:4.1.1}

Classically, the collision of the two domain walls  only occurs at the position of the throat. Semiclassically, the dominant collisions also occur at the point $r_{o}$. The possibility of collision at other positions is small compared to the possibility of  collision occurring at the point $r_{o}$. Thus, for simplicity, we only consider the collision occurring at the point $r_{o}$.   In the later, We will show that the influence of the collision on the tunneling rate can be neglected when $T$ is large enough. This implies that our approximation (ignoring the collisions of the domain wall occurring at other positions) is reasonable.

In the case where the two domain walls collide ``$m$'' times, the classically allowed trajectory of domain wall \uppercase\expandafter{\romannumeral1} is also represented by equation \eqref{eq:4.2}. However, the former ``$m$'' bounces at the point $r_{o}$ are caused by the collision of the two domain walls.   If a bounce is caused by the potential function $V_{2}$ at one time, this implies that when the domain wall  \uppercase\expandafter{\romannumeral1} arrives at the point $r_{o}$, domain wall \uppercase\expandafter{\romannumeral2} has entered the classically allowed region $r_{1}>r_{d}$. Therefore, it is impossible for these two domain walls to collide again.  All subsequent bounces  at the point $r_{o}$ are caused by the potential $V_{2}$. Thus, the ``$m$'' times collisions must correspond to the former ``$m$'' bounces at the point $r_{o}$. After completing the ``$m$'' collisions, the remaining ($N_{1}-m$) bounces at the point $r_{o}$ are caused by the potential $V_{2}$. For those points $r_{o}$ where the bounce is caused by the collision of the two domain walls, one can not insert the I-A-I pair. The reason is that after each collision, the domain wall \uppercase\expandafter{\romannumeral1} will be bounced back into the region $r_{o}<r_{1}<r_{a}$. Consequently, there are $(N_{1}-m)$ points in the trajectory \eqref{eq:4.2} where one can insert the factor $(1+Y)$. All the bounces of the domain wall \uppercase\expandafter{\romannumeral1} at point $r_{a}$ are caused by potential $V_{1}$. Thus, the number of points $r_{a}$ where one can insert the factor $(1+Q)$ is $N_{1}$. Therefore, the amplitude $K^{(m)}(r_{x},\frac{T}{2}; r_{o1}, -\frac{T}{2})$ is
\begin{eqnarray}\begin{split}
\label{eq:4.11}
K^{(m)}(r_{x},\frac{T}{2}; r_{o1}, -\frac{T}{2})=&\triangle_{o1a}e^{-S_{o1a}}\triangle_{oa}^{2(N_{1}-1)}e^{-2(N_{1}-1)S_{oa}}\\&\times\triangle_{ax}e^{-S_{ax}}(1+Q)^{N_{1}}\\&\times(1+Y)^{(N_{1}-m)}.
\end{split}
\end{eqnarray}
When ``$m=0$'', equation \eqref{eq:4.11} becomes \eqref{eq:4.10}.

Substituting equation \eqref{eq:4.11} into \eqref{eq:4.1}, one can obtain the amplitude $K(r_{x},\frac{T}{2}; r_{o1}, -\frac{T}{2})$ as
\begin{eqnarray}\begin{split}
\label{eq:4.14}
K(r_{x},\frac{T}{2}; r_{o1}, -\frac{T}{2})=&\Omega(r_{x})\triangle_{oa}^{2(N_{1}-1)}e^{-2(N_{1}-1)S_{oa}}\\&\times(1+Q)^{N_{1}}(1+Y)^{N_{1}},
\end{split}
\end{eqnarray}
where
\begin{equation}
\label{eq:4.13}
\Omega(r_{x})\equiv (1+Y)Y^{-1}\triangle_{o1a}e^{-S_{o1a}}\triangle_{ax}e^{-S_{ax}}.
\end{equation}
Comparing equations \eqref{eq:4.14} and \eqref{eq:4.10}, one can observe that $K(r_{x},\frac{T}{2}; r_{o1}, -\frac{T}{2})=(1+Y)Y^{-1}K^{(0)}(r_{x},\frac{T}{2}; r_{o1}, -\frac{T}{2})$.  This indicates that the factor $(1+Y)Y^{-1}$ represents the contribution from the collisions.
Equation \eqref{eq:4.14} shows that the collisions, the I-A-I pairs and the classically allowed trajectory are the factors determining the amplitude $K(r_{x},\frac{T}{2}; r_{o1}, -\frac{T}{2})$. The instanton contributed to the amplitude $K(r_{x},\frac{T}{2}; r_{o1}, -\frac{T}{2})$ must be condensed to form a pair with the corresponding anti-instanton.

\subsection{Tunneling rate}
\label{sec:4.2}

The probability of the domain wall \uppercase\expandafter{\romannumeral1} appearing at point $r_{x}$ at the moment $T/2$ is
\begin{eqnarray}\begin{split}
\label{eq:4.15}
P(r_{x},\frac{T}{2}; r_{o1}, -\frac{T}{2})=&\mathscr{N}(T)K(r_{x},\frac{T}{2}; r_{o1}, -\frac{T}{2})\\&\times K^{*}(r_{x},\frac{T}{2}; r_{o1}, -\frac{T}{2}).
\end{split}
\end{eqnarray}
Here, $\mathscr{N}(T)$ represents the normalized constant and $K^{*}(r_{x},\frac{T}{2}; r_{o1}, -\frac{T}{2})$ represents the complex conjugate of $K(r_{x},\frac{T}{2}; r_{o1}, -\frac{T}{2})$. Generally, we denote $X^{*}$ as the complex conjugate of the quantity $X$. If we denote the probability of domain wall \uppercase\expandafter{\romannumeral1} appearing at the classically allowed region $r_{o}<r_{1}<r_{a}$ at the moment $T/2$ as $P(r_{[o,a]},\frac{T}{2}; r_{o1}, -\frac{T}{2})$, then the relationship between $P(r_{[o,a]},\frac{T}{2}; r_{o1}, -\frac{T}{2})$ and $P(r_{x},\frac{T}{2}; r_{o1}, -\frac{T}{2})$ is
\begin{equation}
\label{eq:4.16}
P(r_{[o,a]},\frac{T}{2}; r_{o1}, -\frac{T}{2})=\int_{r_{o}}^{r_{a}}dr_{x}P(r_{x},\frac{T}{2}; r_{o1}, -\frac{T}{2}).
\end{equation}
Substituting equations \eqref{eq:4.14} and \eqref{eq:4.15} into equation \eqref{eq:4.16}, one can show that the probability $P(r_{[o,a]},\frac{T}{2}; r_{o1}, -\frac{T}{2})$ is given by
\begin{eqnarray}\begin{split}
\label{eq:4.17}
P(r_{[o,a]},\frac{T}{2}; r_{o1}, -\frac{T}{2})=&\mathscr{N}(T)\int_{r_{o}}^{r_{a}}dr_{x}\Omega(r_{x})\Omega^{*}(r_{x})\triangle_{oa}^{2(N_{1}-1)}\\&\times(\triangle_{oa}^{2(N_{1}-1)})^{*} e^{-2(N_{1}-1)S_{oa}}\\&\times e^{-2(N_{1}-1)S_{oa}^{*}}
(1+Q)^{N_{1}}(1+Y)^{N_{1}}\\&\times(1+Q^{*})^{N_{1}}(1+Y^{*})^{N_{1}}.
\end{split}
\end{eqnarray}

Defining $\varepsilon_{0}$ as the Euclidean proper time spent by the trajectory $r_{o}\rightarrow r_{a}$, then the factor $(N_{1}-1)$ can be expressed as
\begin{equation}
\label{eq:4.18}
N_{1}-1\approx \frac{T}{2\varepsilon_{0}}\approx N_{1}.
\end{equation}
The approximation in equation \eqref{eq:4.18} is valid in the case where $N_{1}\gg1$ or $T\gg \varepsilon_{0}$.  Substituting equation \eqref{eq:4.18} into \eqref{eq:4.17} and noting that the action $S_{oa}$ is a pure imaginary number (i.e., $S_{oa}^{*}=-S_{oa}$), then the probability $P(r_{[o,a]},\frac{T}{2}; r_{o1}, -\frac{T}{2})$ can be rewritten as
\begin{eqnarray}\begin{split}
\label{eq:4.19}
P(r_{[o,a]},\frac{T}{2}; r_{o1}, -\frac{T}{2})=&\mathscr{N}(T)\int_{r_{o}}^{r_{a}}dr_{x}\Omega(r_{x})\Omega^{*}(r_{x})\\&\times\triangle_{oa}^{\frac{T}{\varepsilon_{0}}}(\triangle_{oa}^{\frac{T}{\varepsilon_{0}}})^{*}\\&\times
(1+Q)^{\frac{T}{2\varepsilon_{0}}}(1+Y)^{\frac{T}{2\varepsilon_{0}}}\\&\times(1+Q^{*})^{\frac{T}{2\varepsilon_{0}}}(1+Y^{*})^{\frac{T}{2\varepsilon_{0}}}.
\end{split}
\end{eqnarray}

The relationship between the tunneling rate $\Gamma$ and the probability $P(r_{[o,a]},\frac{T}{2}; r_{o1}, -\frac{T}{2})$ is~\cite{SC2,RHB}
\begin{equation}
\label{eq:4.20}
P(r_{[o,a]},\frac{T}{2}; r_{o1}, -\frac{T}{2})=e^{-\Gamma T}.
\end{equation}
Combining equations \eqref{eq:4.19} and \eqref{eq:4.20}, one can obtain the tunneling rate as
\begin{eqnarray}\begin{split}
\label{eq:4.21}
\Gamma=&-\frac{1}{T}\mathrm{ln}\int_{r_{o}}^{r_{a}}dr_{x}\Omega(r_{x})\Omega^{*}(r_{x})\\&-\frac{1}{T}\mathrm{ln}\mathscr{N}(T)-\frac{1}{\varepsilon_{0}}\mathrm{ln}\triangle_{oa}\triangle_{oa}^{*}
\\&-\frac{1}{2\varepsilon_{0}}\mathrm{ln}(1+Q)(1+Y)(1+Q^{*})(1+Y^{*}).
\end{split}
\end{eqnarray}
In the case where $T\gg \varepsilon_{0}$, the first term on the right hand side of equation \eqref{eq:4.21} can be neglected. This implies that the contribution of the collisions can be neglected.  Taking the limit of the small fluctuations, the third term on the right hand side of equation \eqref{eq:4.21} can be neglected. When $\triangle_{ab}^{2}\rightarrow 1$ and $\triangle_{oe}^{2}\rightarrow1$ (small fluctuations), according to the definitions \eqref{eq:4.8} and \eqref{eq:4.9}, one can easily show that
\begin{equation}
\label{eq:4.22}
Q\approx \frac{1}{e^{2|S_{ab}|}-1}\approx Q^{*}
\end{equation}
and
\begin{equation}
\label{eq:4.23}
Y\approx \frac{1}{e^{2|S_{oe}|}-1}\approx Y^{*}
\end{equation}
Therefore, in the case of $T\gg \varepsilon_{0}$ and small fluctuations, the tunneling rate $\Gamma$ is given by
\begin{equation}
\label{eq:4.24}
\Gamma=-\frac{1}{T}\mathrm{ln}\mathscr{N}(T)-\frac{1}{\varepsilon_{0}}\mathrm{ln}(1+Q)(1+Y).
\end{equation}

It is difficult to derive the normalized constant $\mathscr{N}(T)$ from the fundamental principles of quantum mechanics. Thus, researchers often avoid discussing the normalized constant when studying the problem of bubble tunneling~\cite{WDJ,SFVF}.  In our model, when the wormhole disappears, the tunneling rate \eqref{eq:4.24} should be equivalent to the result obtained by FGG in~\cite{EAJ}.  This feature of the tunneling rate can help us to infer certain characteristics of the normalized constant $\mathscr{N}(T)$. Specifically, when the wormhole disappears, the only classically forbidden region is $r_{a}<r_{1}<r_{b}$. There is no contribution of the I-A-I pair $r_{o}\rightarrow r_{e}\rightarrow r_{o}$. Thus, the factor $Y$ is zero. The tunneling rate \eqref{eq:4.24} becomes
\begin{eqnarray}\begin{split}
\label{eq:4.25}
\Gamma&\approx-\frac{1}{T}\mathrm{ln}\mathscr{N}(T, r_{o}\rightarrow 0 )-\frac{1}{\varepsilon_{0}}\mathrm{ln}(1+Q(r_{o}\rightarrow0))\\&\approx-\frac{1}{T}\mathrm{ln}\mathscr{N}(T, r_{o}\rightarrow 0 )-\frac{1}{\varepsilon_{0}}e^{-2|S_{ab}(r_{o}\rightarrow 0)|}.
\end{split}
\end{eqnarray}
Here, $\mathscr{N}(T, r_{o}\rightarrow 0 )$, $Q(r_{o}\rightarrow0)$ and $S_{ab}(r_{o}\rightarrow 0)$ represent the normalized constant $\mathscr{N}(T)$, the factor $Q$ and the action $S_{ab}$, respectively, in the limit as the wormhole disappears.

In addition, FGG showed that the tunneling amplitude of the bubble in the spacetime \eqref{eq:1.3} is $e^{-|S_{ab}(r_{o}\rightarrow 0)|}$~\cite{EAJ}. Up to the leading order of the WKB approximation, the tunneling rate is equal to the tunneling probability~\cite{SFVF,EJW}. Thus, the tunneling rate is $e^{-2|S_{ab}(r_{o}\rightarrow 0)|}$. This result should be equivalent to the tunneling rate in equation \eqref{eq:4.25}. This implies that the relationship between $\mathscr{N}(T, r_{o}\rightarrow 0 )$ and $Q(r_{o}\rightarrow 0)$ is
\begin{equation}
\label{eq:4.26}
-\frac{1}{T}\mathrm{ln}\mathscr{N}(T, r_{o}\rightarrow 0 )=\frac{1+\varepsilon_{0}}{\varepsilon_{0}}\mathrm{ln}(1+Q(r_{o}\rightarrow0)).
\end{equation}
Equation \eqref{eq:4.26} shows that the function of the term $-\frac{1}{T}\mathrm{ln}\mathscr{N}(T, r_{o}\rightarrow 0 )$ is to eliminate the factor $-1/\varepsilon_{0}$ in equation \eqref{eq:4.25}. Thus, in order to equivalent to the result obtained by FGG in the limit of the wormhole disappears, the factor $-1/\varepsilon_{0}$ in equation \eqref{eq:4.24} must be eliminated by the first term on the right hand side of equation \eqref{eq:4.24}. Therefore, the most general form of the factor $-\frac{1}{T}\mathrm{ln}\mathscr{N}(T)$ in equation \eqref{eq:4.24} is
\begin{equation}
\label{eq:4.27}
-\frac{1}{T}\mathrm{ln}\mathscr{N}(T)=\frac{1+\varepsilon_{0}}{\varepsilon_{0}}\mathrm{ln}(1+Q)(1+Y)+\mathscr{X}.
\end{equation}
Here, $\mathscr{X}$ represents other terms of the factor $-\frac{1}{T}\mathrm{ln}\mathscr{N}(T)$. The above arguments cannot determine the specific form of $\mathscr{X}$. In the limit of the wormhole disappears, $\mathscr{X}$ must be zero, that is, $\mathscr{X}(r_{o}\rightarrow0)=0$.

Substituting equation \eqref{eq:4.27} into \eqref{eq:4.24}, the tunneling rate $\Gamma$ becomes
\begin{equation}
\label{eq:4.28}
\Gamma=\mathrm{ln}(1+Q)+\mathrm{ln}(1+Y)+\mathscr{X}.
\end{equation}
The tunneling rate $\Gamma$ in equation \eqref{eq:4.28} represents the total tunneling rate from the region $r_{o}<r_{1}<r_{a}$ into the regions $r_{1}>r_{b}$ and $r_{1}>r_{e}$.  In the case where the dominate contribution comes from  single I-A-I trajectories,  we have $\mathrm{ln}(1+Q)\approx e^{-2|S_{ab}|}$ ($\mathrm{ln}(1+Y)\approx e^{-2|S_{oe}|}$).  The factor $e^{-2|S_{ab}|}$ ($e^{-2|S_{oe}|}$) represents the tunneling rate from the region $r_{o}<r_{1}<r_{a}$ into the region $r_{1}>r_{b}$ ($r_{1}>r_{e}$) contributed by a single I-A-I trajectory $r_{a}\rightarrow r_{b}\rightarrow r_{a}$ ($r_{o}\rightarrow r_{e}\rightarrow r_{o}$). This indicates that the factor $\Gamma_{1}\equiv\mathrm{ln}(1+Q)$ ($\Gamma_{2}\equiv\mathrm{ln}(1+Y)$) should be interpreted as the tunneling rate from the region $r_{o}<r_{1}<r_{a}$ into the region $r_{1}>r_{b}$ ($r_{1}>r_{e}$), which includes contributions from both single and multiple I-A-I trajectories.
The domain wall \uppercase\expandafter{\romannumeral1} either tunnels from the region $r_{o}<r_{1}<r_{a}$ into the region $r_{1}>r_{b}$ or tunnels from the region $r_{o}<r_{1}<r_{a}$ into the region $r_{1}>r_{e}$.  Thus, intuitively, $\Gamma_{1}+\Gamma_{2}$ should be equal to the total tunneling rate $\Gamma$. These arguments seems to indicate that $\mathscr{X}=0$. Therefore, the tunneling rate $\Gamma$ is
\begin{equation}
\label{eq:4.29}
\Gamma=\mathrm{ln}(1+Q)(1+Y).
\end{equation}
Equations \eqref{eq:4.22} and \eqref{eq:4.23} show that the factor $Q$ and $Y$ are positive. Thus, the tunneling rate \eqref{eq:4.29} is positive. This implies that the probability of finding the domain wall \uppercase\expandafter{\romannumeral1} in the region $r_{o}<r_{1}<r_{a}$ decreases over time, as it may tunnel into the region $r_{1}>r_{e}$ or $r_{1}>r_{b}$.
One can show that in the limit of the wormhole disappears, the tunneling rate \eqref{eq:4.29} is equivalent to the result obtained by FGG.

\begin{figure}[tbp]
\centering
\includegraphics[width=8.5cm]{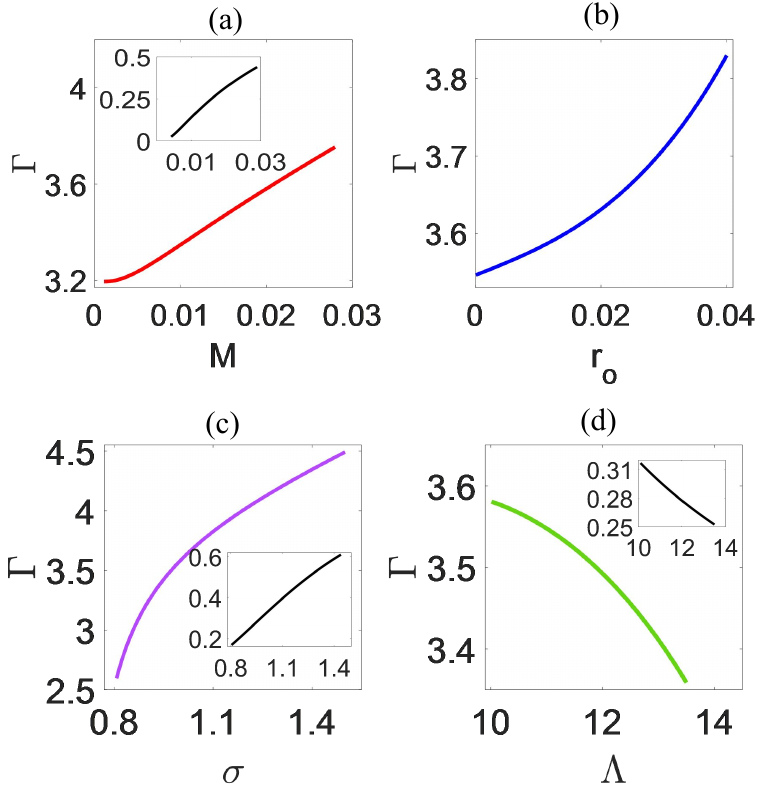}
\caption{\label{fig:10}Variations in the tunneling rate $\Gamma$ with respect to different parameters. In figure (a), the horizontal axis represents the mass parameter of the Schwarzschild spacetime. The parameters are set as $\sigma=1$, $r_{o}=0.01$ and $\Lambda=10$. In figure (b), the horizontal axis corresponds to the radius of the throat, with parameters $\sigma=1$, $M=0.02$ and $\Lambda=10$. In figure (c), the horizontal axis denotes the tension of the domain wall, and the parameters are $M=0.02$, $r_{o}=0.01$ and $\Lambda=10$. In figure (d), the horizontal axis represents the cosmological constant, with $\sigma=1$, $r_{o}=0.01$ and $M=0.02$. The black curves in figures (a), (c) and (d) illustrate the variation of the tunneling rate (vertical axis) with respect to the mass parameter, tension and cosmological constant (horizontal axis), respectively, in the case where the wormhole disappears. The curves in other colors depict the corresponding variations in cases where a wormhole persists. }
\end{figure}

Figure~\ref{fig:10} shows the variations of tunneling rate \eqref{eq:4.29} with respect to certain parameters. The black curves in figures~\ref{fig:10}(a), ~\ref{fig:10}(c) and ~\ref{fig:10}(d) represent the variations of the tunneling rate  over the mass parameter, tension and cosmological constant, respectively, in the case where there is no wormhole. In this case, the tunneling rate is given by $e^{-2|S_{ab}(r_{o}\rightarrow 0)|}$. The curves of other colors represent the variations of the tunneling rate when there exist a wormhole. Figure~\ref{fig:10} shows that the presence of the wormhole enhances the tunneling rate. This result is related to the topological structure of the wormhole. The topology of the throat is $S^{2}\times \mathbb{R}$, and the radius of the throat is nonzero. This structure allows the domain wall to pass though the throat. As a result, both types of single I-A-I trajectories  contribute to the tunneling of the domain wall. Therefore, the number of allowed instanton trajectories increases. Consequently, the topology of the wormhole enhances the tunneling rate. This is similar to the fact that the existence of genus usually increases the number of instantons~\cite{S1}. Our model indicates that the topology of spacetime can influence domain wall tunneling.

Figure~\ref{fig:10}(a) shows that the tunneling rate increases with the mass parameter,  regardless of whether a wormhole is present. Figure~\ref{fig:10}(b) shows that the tunneling rate increases with the radius of the throat. These results can be interpreted intuitively. First, figures~\ref{fig:3}(a) and~\ref{fig:3}(b) show that as the mass parameter increases, the maximal value of the potential $V_{1}$ decreases, and the classically forbidden region $r_{a}<r_{1}<r_{b}$ becomes narrower. It is well known that quantum tunneling usually becomes more difficult as the potential barrier height and the width of the classically forbidden region increase.  This is a typical feature of quantum tunneling. Therefore, a lower potential barrier height and a narrower classically forbidden region correspond to a higher tunneling rate, which is consistent with the result shown in figure~\ref{fig:10}(a). In addition,  as the radius of the wormhole increases, the classically allowed region $r_{o}<r_{1}<r_{a}$ becomes narrower. Consequently, the domain wall has a higher probability of reaching at the tunneling points, which increases the likelihood of transitioning into other classically allowed regions.  This explains why the tunneling rate increases with the throat radius, consistent with the result shown in figure~\ref{fig:10}(b).

Figures~\ref{fig:3}(c) and~\ref{fig:3}(d) show that different values of the surface tension $\sigma$ and cosmological constant $\Lambda$ yield similar shapes of the potential $V_{1}$. According to equation \eqref{eq:2.22}, it is straightforward to verify that the shape of the potential $V_{2}$ is also similar for different values of $\sigma$ and $\Lambda$. This implies that the dynamical behavior of the domain wall remains qualitatively similar across various values of $\sigma$ and $\Lambda$. Thus, for the numerical calculation of the action $S_{ab}$ and the tunneling rate $\Gamma$, the parameters  $\sigma$ and $\Lambda$ are arbitrarily chosen as $\sigma=1$ and $\Lambda=10$, respectively. In this setting, using equations \eqref{eq:2.11} and \eqref{eq:2.12}, the critical mass can be calculated as $M_{cr}\approx 0.0359$. When $M>M_{cr}$, the classical forbidden region $r_{a}<r_{1}<r_{b}$ disappears. Thus, in figures~\ref{fig:7}(a) and~\ref{fig:10}(a), the mass parameter is selected in the range $0<M<0.03$.  In addition, according to equation \eqref{eq:2.10} and the definition of the critical point $r_{a}$, for the case $\sigma=1$, $\Lambda=10$, and $M=0.02$, one obtains $r_{a}\approx 0.0453$. It should be noted that when $r_{o}>r_{a}$,  part of the classical forbidden region $r_{a}<r_{1}<r_{b}$ is removed by the Schwarzschild surgery (and when $r_{o}>r_{b}$, the entire forbidden region $r_{a}<r_{1}<r_{b}$ is removed ). This study focuses on the case where $r_{o}<r_{a}$, such that the region $r_{a}<r_{1}<r_{b}$ is unaffected by the Schwarzschild surgery. Therefore, in figures~\ref{fig:7}(b) and~\ref{fig:10}(b), the value range of the parameter $r_{o}$ is chosen as $0<r_{o}<0.04$.

In our model, there are several independent parameters. When numerically simulating the relationship between the tunneling rate $\Gamma$ and the mass parameter $M$ (or the radius $r_{o}$), the remaining parameters must be held fixed. As such, the quantitative results shown in figures~\ref{fig:10}(a) and~\ref{fig:10}(b) depend on these specific parameter choices. However, we have demonstrated that the qualitative conclusions derived from these figures are consistent with intuitive expectations. These intuitive inferences should not depend on the specific parameter values.  Therefore, we expect the qualitative conclusions presented in  figures~\ref{fig:10}(a) and~\ref{fig:10}(b) to hold independently of particular parameter choices.

Both figures~\ref{fig:10}(a), ~\ref{fig:10}(c), and ~\ref{fig:10}(d) show that the presence of the wormhole enhances the tunneling rate.  Within the parameter range shown in figures~\ref{fig:10}(c) and ~\ref{fig:10}(d), the tunneling rate increases with the surface tension $\sigma$, while it decreases with the cosmological constant $\Lambda$, regardless of whether a wormhole exists or not.  When the wormhole disappears, our model reduces to the one studied by FGG in~\cite{EAJ}. Without the modification of the terms related to the Heaviside function, FGG demonstrated that the action $S_{ab}$  in equation \eqref{eq:3.19ma} is not a continuous function of the mass $M$. Using the same methods, one can show that the action $S_{ab}$ described by equation \eqref{eq:3.19} or \eqref{eq:3.19ma} is also not a continuous function of  $\sigma$ or $\Lambda$ over a broader parameter range. This issue exists both in our model and in the model studied by FGG in~\cite{EAJ}. To our knowledge, this issue remains unsolved. Thus, it is not straightforward to study the variation of the tunneling rate with respect to $\sigma$ or $\Lambda$ over the whole parameter space. Nevertheless, since the main qualitative conclusions obtained by FGG in~\cite{EAJ} are consistent with those presented by Fischer, Morgan and Polchinski using a different method in~\cite{WDJ} (see the comments in~\cite{SFVF}), we expect that this issue does not prevent us from qualitatively analyzing the variation of the tunneling rate with respect to the mass $M$ or the radius of the throat $r_{o}$.

\begin{figure}[tbp]
\centering
\includegraphics[width=8.5cm]{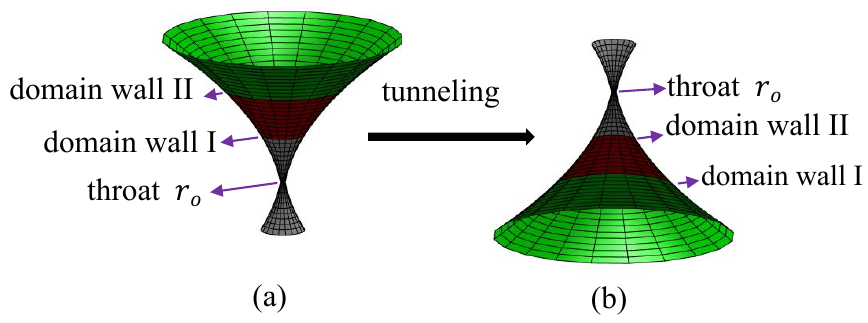}
\caption{\label{fig:11} The bubble tunnels from one universe to another. The green sections represent Schwarzschild spacetime. The red sections represent the bubble (dS spacetime). The grey sections represent Minkowski spacetime. A different side of the throat is defined as a different universe. Figure (a) and (b) depict the spacetime structure before and after the tunneling, respectively. }
\end{figure}

To derive equation \eqref{eq:4.29}, we made the assumption that in the initial state, the two domain walls are positioned on opposite sides of the throat and have equal radii. Thus, the dominant collisions between the domain walls  take place at the throat. This assumption simplifies the derivation process. However, we have shown that the impact of the collisions on the tunneling rate can be neglected in the case of $T\gg\varepsilon_{0}$. We expect this conclusion be still valid in the case where the dominant collision occurs at other points.  Setting $K^{(m)}(r_{x},\frac{T}{2}; r_{o1}, -\frac{T}{2})=0$ (neglecting the effect of collision), the above derivations are still valid for the case where the two domain walls have different radii in the initial state  (we also need to require that domain wall \uppercase\expandafter{\romannumeral2} does not pass through the throat during the evolution process). Thus, one can infer that equation \eqref{eq:4.29} is valid regardless of whether the radii of the two domain walls are the same or not in the initial state.

For the case where both domain walls are located at the same side of the throat at the initial time, as depicted in figure~\ref{fig:1}(d), equation \eqref{eq:4.29} is not valid. Nevertheless, one can still  employ the analogous approaches presented in this section to study the tunneling of the domain wall. Assuming that in the initial state, $r_{e}<r_{o1}<r_{o2}$, $r_{o2}>r_{d}$, $\dot{r}_{o1}<0$ and $\dot{r}_{o2}<0$. That is, both of these two domain walls are contracting. Before passing through the throat, the dynamics of domain wall \uppercase\expandafter{\romannumeral1} are described by equation \eqref{eq:2.18}, while the dynamics of domain wall \uppercase\expandafter{\romannumeral2} are described by equation \eqref{eq:2.8}. In the case where the dominate contribution comes from single I-A-I trajecties, the amplitude of domain wall \uppercase\expandafter{\romannumeral1} (\uppercase\expandafter{\romannumeral2}) passing through the classically forbidden region $r_{o}<r_{1}<r_{e}$ ($r_{c}<r_{1}<r_{d}$) is $e^{-|S_{oe}|}$ ($e^{-|S_{ab}|}$).

Subsequently, the domain wall \uppercase\expandafter{\romannumeral1} will go across the throat and then enter the classically allowed region $r_{o}<r_{1}<r_{a}$. For the domain wall \uppercase\expandafter{\romannumeral2}, after it passes through the classically forbidden region $r_{c}<r_{1}<r_{d}$ via quantum tunneling, it will enter the classically allowed region $r_{o}<r_{1}<r_{c}$. Subsequently, it may go across the throat and pass through the classically forbidden region $r_{o}<r_{1}<r_{e}$ with the tunneling amplitude $e^{-|S_{oe}|}$. Thus, the amplitude of domain wall \uppercase\expandafter{\romannumeral2} entering the other side of the throat is  $e^{-|S_{ab}|}e^{-|S_{oe}|}$. Therefore, the tunneling probability (tunneling rate) for both of these two domain walls across the throat is
\begin{equation}
\label{eq:4.30}
\Gamma=e^{-2|S_{oe}|}e^{-2|S_{ab}|}e^{-2|S_{oe}|}.
\end{equation}
Equation \eqref{eq:4.30} does not account for the influence of multiple instanton trajectories.  One can define the different sides of the throat as different universes. Assuming that within a universe, there exists a bubble and within that bubble, there is a wormhole which is connected to another universe (figure~\ref{fig:1}(d)). These arguments suggest that the bubble can go across the throat and enter another universe through the quantum tunneling, as illustrated in figure~\ref{fig:11}, even though it is prohibited by the classical physics.

To sum up, in this section, we have derived the tunneling rate of the bubble, which is described by equation \eqref{eq:4.29}. It incorporates the contributions from both single and multiple I-A-I trajectories. The results in figure~\ref{fig:10} are consistent with intuitive expectations, which indicates that formula \eqref{eq:4.29} is reasonable in some sense. Moreover, the effect of the conical singularity at the horizon has not been taken into account in figure~\ref{fig:10}. A detailed discussion about the conical singularity is provided in appendix~\ref{sec:E}.

\section{Conclusions and discussions}
\label{sec:5}

In this work, we use Schwarzschild surgery to construct a wormhole.  We introduce certain matters to glue two manifolds into a spacetime. We have neglected the interaction between the bubble and the matters, as well as the dynamics of the throat. Usually, the topology of the wormhole is non-trivial. Thus our model may help to reveal the influence of the topology on the false vacuum bubble tunneling.

In our model, the bubble has two domain walls, we assume that at the initial time, the two domain walls are located at opposite sides of the throat.  Inside the domain wall is the dS spacetime, and outside the domain wall is the Schwarzschild spacetime. The junction condition of the extrinsic curvature determines the dynamics of the domain wall. The two domain walls may collide at the position of the throat. For simplicity, we assume that the collision is elastic. The dynamics of the two domain walls are identical. Thus, we only focus on the dynamics of one of the domain walls. For each domain wall, there are two classically forbidden regions and three classically allowed regions if the effective  energy of the domain wall is smaller than the critical mass. When the domain wall passes through the throat, its effective potential will change. And the spacetime inside the domain wall will become Minkowski spacetime.

We have derived the formula for the tunneling rate. We sum all the trajectories that contribute to the tunneling rate. There are two types of single I-A-I trajectories that contribute to the tunneling of the domain wall. The topology of the throat is $S^{2}\times \mathbb{R}$, and its radius is nonzero. This structure allows the domain wall to pass though the throat. Consequently, the two types of single I-A-I trajectories can combine to form an infinite number of multiple I-A-I trajectories, all of which contribute to tunneling. The increase in  the number of instanton trajectories enhances the tunneling rate.  Thus, the topology of the wormhole leads to an increase in the tunneling rate. Our formula for the tunneling rate includes contributions from both single and multiple I-A-I trajectories.  In the limit where the wormhole disappears, there is only one classically forbidden region, implying that there is only one type of single I-A-I trajectory.

Finally, we demonstrate that in the limit of a long time, the effect of the domain wall collision can be neglected. We find that as the increases of the radius of the throat, the tunneling rate also increases. We also find that an increase in the mass of the black hole  enhances the tunneling rate.  We show that if there exists a wormhole within the bubble, and the wormhole is connected to another universe, then it is possible that this bubble enters another universe through quantum tunneling.

\section*{Acknowledgements}
Hong Wang  was supported by the National Natural Science Foundation of China Grant  No.12234019. Hong Wang thanks for the help from Professor Erkang Wang.

\appendix

\section{Representation of Schwarzschild surgery in the Penrose diagram}
\label{sec:A1}

\begin{figure}[tbp]
\centering
\includegraphics[width=6cm]{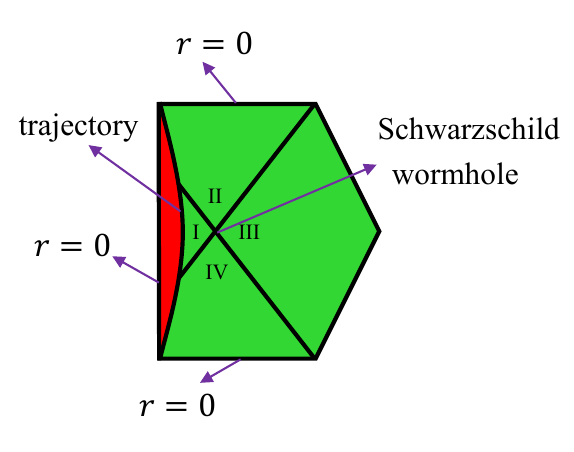}
\caption{\label{fig:1a} Penrose diagram of the spacetime defined by equation \eqref{eq:1.3}. The red part and green part represent the dS spacetime and the Schwarzschild spacetime, respectively. The boundary between the red and green parts represents the trajectory of the domain wall. }
\end{figure}

\begin{figure*}[tbp]
\centering
\includegraphics[width=11cm]{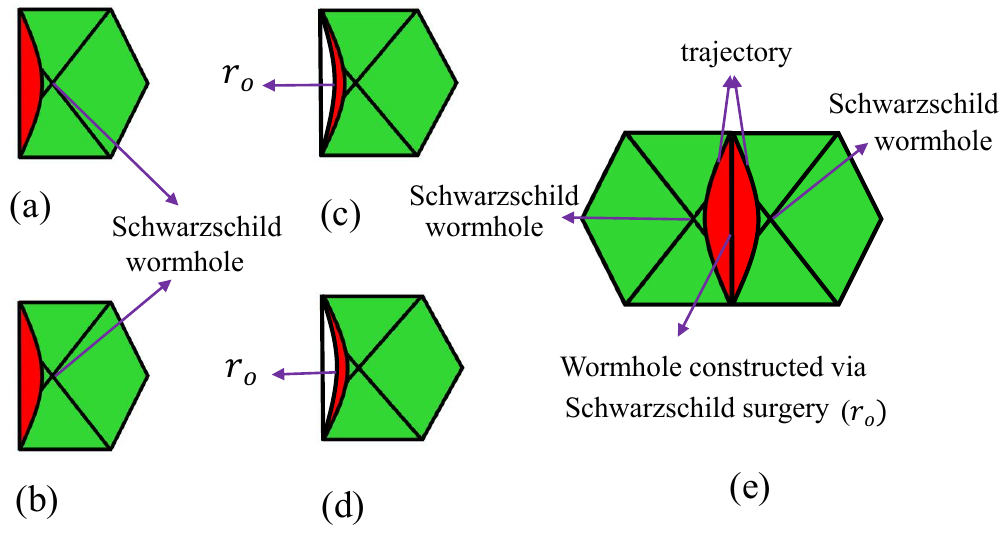}
\caption{\label{fig:1b} Penrose diagram for the Schwarzschild surgery. The red parts and the green parts in figures (a)-(e) represent the dS spacetime and the Schwarschild spacetime, respectively. Figures (a) and (b) represent the manifolds defined by equation \eqref{eq:1.3}. The white parts in figures (c) and (d) are removed in the Schwarzschild surgery. By pasting the manifolds in figures (c) and (d) along the boundary $r_{o}$, one can obtain the manifold depicted in figure (e).  }
\end{figure*}

It is useful to represent the Schwarzschild surgery via the Penrose diagram. Figure~\ref{fig:1a} represents the Penrose diagram of the spacetime defined by equation \eqref{eq:1.3}~\cite{SFVF}.  Various allowed trajectories exist for the domain wall, with each trajectory corresponding to a different Penrose diagram, all of which are similar (inside the bubble is dS spacetime, and outside the bubble is Schwarzschild spacetime). For convenience, we focus on the Penrose diagram where the coordinate radius of the domain wall increases from zero to a maximum value beyond the Schwarzschild radius, and then collapses back to zero,  as shown in figure~\ref{fig:1a}. The Penrose diagram of the Schwarzschild surgery for the spacetime defined by equation \eqref{eq:1.3} is shown in figure~\ref{fig:1b}.  Figures~\ref{fig:1b}(a) and ~\ref{fig:1b}(b) represent the spacetimes defined by equation \eqref{eq:1.3}. The white regions in figures~\ref{fig:1b}(c) and ~\ref{fig:1b}(d) are removed during Schwarzschild surgery. By gluing the manifolds in figures~\ref{fig:1b}(c) and ~\ref{fig:1b}(d) along the boundary $r_{o}$, one obtains the manifold in figure~\ref{fig:1b}(e), which is the Penrose diagram of the manifold in figure~\ref{fig:1}(c).

By comparing  figures~\ref{fig:1a} and~\ref{fig:1b}(e), one can observe the differences between the bubble studied by FGG in~\cite{EAJ} and the bubble depicted in figure~\ref{fig:1}(c). In figure~\ref{fig:1a}, there is a wormhole connecting regions \uppercase\expandafter{\romannumeral1} and \uppercase\expandafter{\romannumeral3}~\cite{SEA}. We refer to this as the Schwarzschild wormhole. However, in figure~\ref{fig:1b}(e), in addition to the wormhole constructed via Schwarzschild surgery (with a coordinate radius $r_{o}$ of the throat), there are two additional Schwarzschild wormholes (with a coordinate radius $2M$ of the throats ). The tunneling of the bubble in figure~\ref{fig:1a} is fully determined by the dynamical equation of the domain wall~\cite{EAJ}.  Although the domain wall may pass through the Schwarzschild wormhole, it does not affect the tunneling of the bubble~\cite{EAJ}. Furthermore, the affect of the conical singularity is discussed in appendix~\ref{sec:E}.  In contrast, as we  have shown  in sections~\ref{sec:3} and~\ref{sec:4}, the wormhole constructed via Schwarzschild surgery can influence the tunneling of the bubble.

\section{Derivation of equation \eqref{eq:2.8}}
\label{sec:A1a}

Starting from the Einstein equations, one can derive the dynamical equation of the domain wall in Gaussian normal coordinates as~\cite{SEA}
\begin{equation}
\label{eq:2.2}
\lim_{\varepsilon\rightarrow 0}[K_{ab}(\eta=+\epsilon)-K_{ab}(\eta=-\epsilon)]=-8\pi(\mathfrak{S}_{ab}-\frac{1}{2}h_{ab}\mathfrak{S}).
\end{equation}
Here, $\eta$ is the proper distance away from the domain wall, $K_{ab}$ is the extrinsic curvature,  $h_{ab}$ is the induced metric on the domain wall and $\mathfrak{S}_{ab}$ is the  energy momentum tensor of the domain wall. $\mathfrak{S}$ represents the trace of $\mathfrak{S}_{ab}$, that is, $\mathfrak{S}=h^{ab}\mathfrak{S}_{ab}$. Under the thin wall approximation, one can show that the relationship between $\mathfrak{S}_{ab}$ and $h_{ab}$ is~\cite{SEA}
\begin{equation}
\label{eq:2.3}
\mathfrak{S}_{ab}=-\sigma h_{ab}.
\end{equation}
For the bubble as presented in figure~\ref{fig:1}(c), inside the domain wall is the dS spacetime and outside the domain wall  is the Schwarzschild spacetime. Thus, the 4-velocity (time-like vector) of the domain wall can be expressed as $U_{s}^{\mu}=(\dot{t}_{s}, \dot{r}_{1}, 0, 0)$ or $U_{D}^{\mu}=(\dot{t}_{D}, \dot{r}_{1}, 0, 0)$. Here, $\dot{t}_{s}=dt_{s}/d\tau$, $\dot{t}_{D}=dt_{D}/d\tau$ and $\dot{r}_{1}=dr_{1}/d\tau$.  The related unit normal vectors (space-like) are $\xi_{ s,  \mu}=(-\dot{r}_{1}, \dot{t}_{s}, 0, 0)$ and $\xi_{ D,  \mu}=(\dot{r}_{1}, -\dot{t}_{D}, 0, 0)$, respectively. One can easily prove that $U_{s}^{\mu}\xi_{ s,  \mu}=0$ and $U_{D}^{\nu}\xi_{ D,  \nu}=0$ ($\mu, \nu=0, 1, 2, 3$). Noted that $\dot{r}_{1}$ is the radial velocity of the domain wall.

The normalization condition for $U_{s}^{\mu}$ is $g_{s,\mu\nu}U_{s}^{\mu}U_{s}^{\nu}=-1$. We use $g_{s,\mu\nu}$ to represent the metric of the Schwarzschild spacetime, $g_{s,00}=-A_{s}$ and $g_{s,11}=A_{s}^{-1}$. Combining this normalization condition with equations \eqref{eq:1.1} and \eqref{eq:1.2}, one can obtain~\cite{RJ}
\begin{equation}
\label{eq:2.4}
A_{s}\dot{t}^{2}_{s}-A_{s}^{-1}\dot{r}_{1}^{2}=1.
\end{equation}
Similarly, the normalization condition for $U_{D}^{\mu}$ is $g_{D,\mu\nu}U_{D}^{\mu}U_{D}^{\nu}=-1$ where $g_{D,\mu\nu}$ represents the metric of the dS spacetime, $g_{D,00}=-A_{D}$ and $g_{D,11}=A_{D}^{-1}$. Combining equations \eqref{eq:1.3}, \eqref{eq:1.4} and the normalization condition for $U_{D}^{\mu}$, one can obtain
\begin{equation}
\label{eq:2.5}
A_{D}\dot{t}^{2}_{D}-A_{D}^{-1}\dot{r}_{1}^{2}=1.
\end{equation}

Noted that $\lim_{\varepsilon\rightarrow 0}K_{ab}(\eta=-\epsilon)=\nabla_{a}\xi_{ D,b}$ and $\lim_{\varepsilon\rightarrow 0}K_{ab}(\eta=+\epsilon)=\nabla_{a}\xi_{ s,b}$. Combining the definition of the extrinsic curvature with equations \eqref{eq:1.2}, \eqref{eq:1.3} and \eqref{eq:1.4}, one can obtain~\cite{RJ}
\begin{equation}
\label{eq:2.6}
\lim_{\varepsilon\rightarrow 0}K_{11}(\eta=+\epsilon)=A_{s}r_{1}\dot{t}_{s}
\end{equation}
and
\begin{equation}
\label{eq:2.7}
\lim_{\varepsilon\rightarrow 0}K_{11}(\eta=-\epsilon)=-A_{D}r_{1}\dot{t}_{D}.
\end{equation}
Bringing equation \eqref{eq:1.5} into the right hand side of equation \eqref{eq:2.2}, we have $-8\pi(\mathfrak{S}_{11}-\frac{1}{2}h_{11}\mathfrak{S})=-4\pi\sigma r_{1}^{2}$. Combining this formula with equations \eqref{eq:2.2}, \eqref{eq:2.6} and \eqref{eq:2.7}, one can obtain the dynamical equation \eqref{eq:2.8}.

\section{Derivation of equation \eqref{eq:3.19}}
\label{sec:A2}

Combining equations \eqref{eq:3.4}-\eqref{eq:3.9}, one can show that the action $S_{ab}$ can be written as
\begin{equation}
\label{eq:3.9a}
S_{ab}=I^{tot}_{ab,E}\Big|_{r_{o}\rightarrow 0}-I^{sub}_{ab,E}\Big|_{r_{o}\rightarrow 0}+I_{ab,E}\Big|_{r_{o}}-I^{sub}_{ab,E}\Big|_{r_{o}}.
\end{equation}
Here,
\begin{eqnarray}\begin{split}
\label{eq:3.9b}
I^{tot}_{ab,E}\Big|_{r_{o}\rightarrow 0}\equiv&4\pi\sigma \int_{\tau_{a}}^{\tau_{b}}r_{1}^{2}d\tau_{E}-\frac{1}{16\pi}\\&\times\lim_{\epsilon\rightarrow0}\int_{-\epsilon}^{+\epsilon}d\eta\int_{\tau_{a}}^{\tau_{b}}d\tau_{E}\int d\theta d\phi r^{2}sin \theta\mathscr{R}\\&-\int_{t_{DEa}}^{t_{DEb}}dt_{DE}\int_{0}^{r_{1}}dr\int d\theta d\phi r^{2}sin \theta (\Lambda+\frac{\mathscr{R}}{16\pi}),
\end{split}
\end{eqnarray}
\begin{equation}
\label{eq:3.9c}
I_{ab,E}\Big|_{r_{o}}\equiv \int_{t_{DEa}}^{t_{DEb}}dt_{DE}\int_{0}^{r_{o}}dr\int d\theta d\phi r^{2}sin \theta (\Lambda+\frac{\mathscr{R}}{16\pi}).
\end{equation}
In equation \eqref{eq:3.9a}, $I^{sub}_{ab,E}\Big|_{r_{o}\rightarrow 0}$ and $I^{sub}_{ab,E}\Big|_{r_{o}}$ are the subtracted term correspond to $I^{tot}_{ab,E}\Big|_{r_{o}\rightarrow 0}$ and $I_{ab,E}\Big|_{r_{o}}$, respectively. It is straightforward to verify that $I^{tot}_{ab,E}=I^{tot}_{ab,E}\Big|_{r_{o}\rightarrow 0}+I_{ab,E}\Big|_{r_{o}}$ and $I^{sub}_{ab,E}=I^{sub}_{ab,E}\Big|_{r_{o}\rightarrow 0}+I^{sub}_{ab,E}\Big|_{r_{o}}$. The term $I^{tot}_{ab,E}\Big|_{r_{o}\rightarrow 0}-I^{sub}_{ab,E}\Big|_{r_{o}\rightarrow 0}$ represents the subtracted tunneling action when the wormhole disappears. The term $I_{ab,E}\Big|_{r_{o}}-I^{sub}_{ab,E}\Big|_{r_{o}}$ represents the contribution of the wormhole to the action $S_{ab}$.

In~\cite{EAJ}, FGG proved that
\begin{eqnarray}\begin{split}
\label{eq:3.9d}
I^{tot}_{ab,E}\Big|_{r_{o}\rightarrow 0}-I^{sub}_{ab,E}\Big|_{r_{o}\rightarrow 0}=&-2\pi\sigma \int_{\tau_{a}}^{\tau_{b}}r_{1}^{2}d\tau_{E}\\&-\frac{4}{3}\pi\Lambda\int_{\tau_{a}}^{\tau_{b}}d\tau_{E}\frac{\beta_{D1}}{A_{D}}r_{1}^{3}
\\&-\frac{1}{2}M\Big\{\int_{\tau_{a}}^{\tau_{b}}\frac{\beta_{s}}{A_{s}}d\tau_{E}\\&-4\pi M\Theta(M_{s}-M)\Big\}\\&-\frac{4}{3}\pi^{2}\Lambda\chi^{-4}\Theta(M_{D}-M).
\end{split}
\end{eqnarray}
Substituting $\mathscr{R}=-32\pi\Lambda$ into equation \eqref{eq:3.9c}, and performing the integration over the variables ($r, \theta, \phi$), equation \eqref{eq:3.9c} becomes
\begin{eqnarray}\begin{split}
\label{eq:A2a}
I_{ab,E}\Big|_{r_{o}}&=-\frac{4}{3}\pi\Lambda r_{o}^{3}\int_{t_{DEa}}^{t_{DEb}}dt_{DE}\\&=\frac{4}{3}\pi\Lambda r_{o}^{3}\int_{\tau_{a}}^{\tau_{b}}d\tau_{E}\frac{\beta_{D1}}{A_{D}}.
\end{split}
\end{eqnarray}
From the first to the second step, we used the definition $\beta_{D1}\equiv -A_{D}\dot{t}_{D}$. According to the definition of the subtracted term,  the term $I_{ab,E}^{sub}\Big|_{r_{o}}$ can be written as
\begin{eqnarray}\begin{split}
\label{eq:A2b}
I_{ab,E}^{sub}\Big|_{r_{o}}=\frac{4}{3}\pi\Lambda r_{o}^{3}\int_{\tau_{a}}^{\tau_{b}}d\tau_{E}\frac{\beta_{D1}}{A_{D}}\Big|_{\dot{r}_{1}=0}.
\end{split}
\end{eqnarray}
Noting that at the critical point, $A_{D}=\beta_{D1}^{2}$. Then, equation \eqref{eq:A2b} can be simplified as
\begin{equation}
\label{eq:A2c}
I_{ab,E}^{sub}\Big|_{r_{o}}=\frac{4}{3}\pi\Lambda r_{o}^{3}\beta_{D1}^{-1}\Big|_{\dot{r}_{1}=0}\int_{\tau_{a}}^{\tau_{b}}d\tau_{E}.
\end{equation}

Combining equations \eqref{eq:A2a} and \eqref{eq:A2c}, one obtains
\begin{eqnarray}\begin{split}
\label{eq:A2d}
I_{ab,E}\Big|_{r_{o}}-I_{ab,E}^{sub}\Big|_{r_{o}}=&\frac{4}{3}\pi\Lambda r_{o}^{3}\int_{\tau_{a}}^{\tau_{b}}d\tau_{E}\frac{\beta_{D1}}{A_{D}}\\&-\frac{4}{3}\pi\Lambda r_{o}^{3}\beta_{D1}^{-1}\Big|_{\dot{r}_{1}=0}\int_{\tau_{a}}^{\tau_{b}}d\tau_{E}.
\end{split}
\end{eqnarray}
However, this equation is only valid in the case where $M>M_{D}$. For a more general case, equation \eqref{eq:A2d} should be modified to
\begin{eqnarray}\begin{split}
\label{eq:A2e}%
I_{ab,E}\Big|_{r_{o}}-I_{ab,E}^{sub}\Big|_{r_{o}}=&\frac{4}{3}\pi\Lambda r_{o}^{3}\int_{\tau_{a}}^{\tau_{b}}d\tau_{E}\frac{\beta_{D1}}{A_{D}}\\&+\frac{4}{3}\pi^{2}\Lambda\chi^{-1}r_{o}^{3}\Theta(M_{D}-M)\\&-\frac{4}{3}\pi\Lambda r_{o}^{3}\beta_{D1}^{-1}\Big|_{\dot{r}_{1}=0}\int_{\tau_{a}}^{\tau_{b}}d\tau_{E}.
\end{split}
\end{eqnarray}
Substituting equations \eqref{eq:3.9d} and \eqref{eq:A2e} into  equation \eqref{eq:3.9a}, one obtains equation \eqref{eq:3.19}.

To prove equation \eqref{eq:A2e},  we first consider the case where the wormhole disappears, the bubble only has one domain wall. Inside the domain wall is  dS spacetime, and outside the domain wall is Schwarzschild spacetime. In the case of $M>M_{D}$, $\beta_{D1}$ is always positive in the classically forbidden region $r_{a}<r_{1}<r_{b}$. The factor
\begin{equation}
\label{eq:a1}
\frac{4}{3}\pi\int_{\tau_{a}}^{\tau_{b}}d\tau_{E}\frac{\beta_{D1}}{A_{D}}r_{1}^{3}
\end{equation}
represents the four-volume swept out by the bubble when it tunnels from $r_{a}$ to $r_{b}$. In equation \eqref{eq:a1}, $\frac{4}{3}\pi r_{1}^{3}$ represents the comoving volume of the bubble, and $\int_{\tau_{a}}^{\tau_{b}}d\tau_{E}\frac{\beta_{D1}}{A_{D}}$ represents the time spent by the bubble when it tunnels from $r_{a}$ to $r_{b}$. In the case of $M<M_{D}$, it is
\begin{equation}
\label{eq:a2}
\frac{4}{3}\pi\int_{\tau_{a}}^{\tau_{b}}d\tau_{E}\frac{\beta_{D1}}{A_{D}}r_{1}^{3}+\frac{4}{3}\pi\chi^{-3}\cdot\pi\chi^{-1}
\end{equation}
represents the four-volume swept out by the bubble when it tunnels from $r_{a}$ to $r_{b}$. The second term in equation \eqref{eq:a2} is half of the four-volume of the Euclidean dS spacetime. The factor $\frac{4}{3}\pi\chi^{-3}$ represents the comoving volume of the dS spacetime. The factor $\pi\chi^{-1}$ is the time parameter. Thus, in the more general case, the factor \eqref{eq:a1} should be modified as
\begin{equation}
\label{eq:A2f}
\frac{4}{3}\pi\int_{\tau_{a}}^{\tau_{b}}d\tau_{E}\frac{\beta_{D1}}{A_{D}}r_{1}^{3}+\frac{4}{3}\pi\chi^{-3}\cdot\pi\chi^{-1}\Theta(M_{D}-M).
\end{equation}
Detailed derivations of equations \eqref{eq:a2} and \eqref{eq:A2f} can be found in~\cite{EAJ}.

In our model, there is a wormhole. The bubble has two domain walls. On one side of the throat, the comoving volume of the bubble is $\frac{4}{3}\pi (r_{1}^{3}-r_{o}^{3})$. Noted that the Schwarzschild surgery does not change the time parameter $\int_{\tau_{a}}^{\tau_{b}}d\tau_{E}\frac{\beta_{D1}}{A_{D}}$ and $\pi\chi^{-1}$. Thus, the first term in equation \eqref{eq:A2f} should be replaced by
\begin{equation}
\label{eq:a3}
\frac{4}{3}\pi\int_{\tau_{a}}^{\tau_{b}}d\tau_{E}\frac{\beta_{D1}}{A_{D}}(r_{1}^{3}-r_{o}^{3}).
\end{equation}
And the second term in equation \eqref{eq:A2f} should be replaced by
\begin{equation}
\label{eq:a4}
(\frac{4}{3}\pi\chi^{-3}-\frac{4}{3}\pi r_{o}^{3})\cdot\pi\chi^{-1}\Theta(M_{D}-M).
\end{equation}
Here, we have subtracted the factor $\frac{4}{3}\pi r_{o}^{3}$ as the region where $r<r_{o}$ is removed in the Schwarzschild surgery. Therefore, in the general case, equation \eqref{eq:a3} should be modified to
\begin{equation}
\label{eq:a5}
\frac{4}{3}\pi\int_{\tau_{a}}^{\tau_{b}}d\tau_{E}\frac{\beta_{D1}}{A_{D}}(r_{1}^{3}-r_{o}^{3})+\frac{4}{3}\pi(\chi^{-3}- r_{o}^{3})\pi\chi^{-1}\Theta(M_{D}-M).
\end{equation}
Combining equations \eqref{eq:a1}, \eqref{eq:A2f}, \eqref{eq:a3} and \eqref{eq:a5}, one can show that  the factor $\frac{4}{3}\pi r_{o}^{3}\int_{\tau_{a}}^{\tau_{b}}d\tau_{E}\frac{\beta_{D1}}{A_{D}}$ in equation \eqref{eq:A2d} should be modified to
\begin{equation}
\label{eq:a6}
\frac{4}{3}\pi r_{o}^{3}\int_{\tau_{a}}^{\tau_{b}}d\tau_{E}\frac{\beta_{D1}}{A_{D}}+\frac{4}{3}\pi r_{o}^{3}\pi\chi^{-1}\Theta(M_{D}-M)
\end{equation}
in the general case. From equations \eqref{eq:A2d} and \eqref{eq:a6}, one can obtain equation \eqref{eq:A2e}.

\section{Derivation of equation \eqref{eq:3.27} }
\label{sec:B}

Substituting equations \eqref{eq:3.21}-\eqref{eq:3.24} into equation \eqref{eq:3.20}, and performing the integration over the variables $(r, \theta, \phi)$, one can show that action $S_{oe}$ is given by
\begin{eqnarray}\begin{split}
\label{eq:3.24a}%
S_{oe}=&-\frac{4}{3}\pi\Lambda\int_{\tau_{o}}^{\tau_{e}}r_{o}^{3}\frac{\beta_{D2}}{A_{D}}d\tau_{E}+8\pi\sigma\int_{\tau_{o}}^{\tau_{e}}r_{1}^{2}d\tau_{E}\\&-\int_{\tau_{o}}^{\tau_{e}}L_{0oe}d\tau_{E}-I^{sub}_{oe,E}
\end{split}
\end{eqnarray}
where
\begin{eqnarray}\begin{split}
\label{eq:3.24b}
L_{0oe}(r_{1}, \dot{r}_{1})\equiv & 4\pi\sigma r_{1}^{2}-\frac{4}{3}\pi\Lambda\frac{\beta_{D2}}{A_{D}}r_{1}^{3}
+r_{1}(\beta_{M}-\beta_{D2})\\&+\frac{1}{2}r_{1}^{2}\frac{d\dot{r}_{1}}{d\tau_{E}}(\frac{1}{\beta_{D2}}-\frac{1}{\beta_{M}})\\&-
\frac{1}{4}r_{1}^{2}\frac{1}{\beta_{D2}}\frac{dA_{D}}{dr_{1}}.
\end{split}
\end{eqnarray}
In addition, FGG proved that~\cite{EAJ}
\begin{eqnarray}\begin{split}
\label{eq:3.24mc}%
I^{tot}_{ab,E}=\frac{4}{3}\pi\Lambda\int_{\tau_{a}}^{\tau_{b}}d\tau_{E}\frac{\beta_{D1}}{A_{D}}r_{o}^{3}+ \int_{\tau_{a}}^{\tau_{b}}L_{0ab}d\tau_{E},
\end{split}
\end{eqnarray}
where
\begin{eqnarray}\begin{split}
\label{eq:3.24md}
L_{0ab}(r_{1},\dot{r}_{1})\equiv &4\pi\sigma r_{1}^{2}-\frac{4}{3}\pi\Lambda\frac{\beta_{D1}}{A_{D}}r_{1}^{3}
+r_{1}(\beta_{s}-\beta_{D1})\\&+\frac{1}{2}r_{1}^{2}\frac{d\dot{r}_{1}}{d\tau_{E}}(\frac{1}{\beta_{D1}}-\frac{1}{\beta_{s}})\\&-
\frac{1}{4}r_{1}^{2}(\frac{1}{\beta_{D1}}\frac{dA_{D}}{dr_{1}}-\frac{1}{\beta_{s}}\frac{dA_{s}}{dr_{1}}).
\end{split}
\end{eqnarray}
Equations \eqref{eq:3.24md} is consistent with equation (3.17) in reference~\cite{EAJ}.

Noted that when $M\rightarrow 0$, the dynamical equation \eqref{eq:2.9} becomes \eqref{eq:2.21}, $\beta_{s}\rightarrow \beta_{M}$, $\beta_{D1}\rightarrow \beta_{D2}$ and $\partial_{r_{1}}A_{s}\Big|_{M\rightarrow0}=0$. Thus, $L_{0ab}(r_{1}, \dot{r}_{1})\Big|_{M\rightarrow 0}=L_{0oe}(r_{1}, \dot{r}_{1})$. In~\cite{EAJ}, FGG shown that $\int_{\tau_{a}}^{\tau_{b}}L_{0ab}\Big|_{\dot{r}_{1}=0}d\tau_{E}=\frac{1}{2}M\mathscr{T}_{\infty}$. Here, $\mathscr{T}_{\infty}$ represents the Euclidean Schwarzschild coordinate time measured at infinity. Combining these equalities,  one can show that
\begin{eqnarray}\begin{split}
\label{eq:3.24c}%
L_{0oe}(r_{1}&,\dot{r}_{1}=0)\int_{\tau_{o}}^{\tau_{e}}d\tau_{E}\\&\propto L_{0ab}(r_{1},\dot{r}_{1}=0)\Big|_{M\rightarrow 0}\int_{\tau_{a}}^{\tau_{b}}d\tau_{E}=0.
\end{split}
\end{eqnarray}
Therefore, the subtracted term $I^{sub}_{oe,E}$ is
\begin{equation}
\label{eq:3.24d}%
I^{sub}_{oe,E}=-\frac{4}{3}\pi\Lambda r_{o}^{3}\frac{\beta_{D2}}{A_{D}}\Big|_{\dot{r}_{1}=0}\int_{\tau_{o}}^{\tau_{e}}d\tau_{E}+8\pi\sigma r_{e}^{2}\int_{\tau_{o}}^{\tau_{e}}d\tau_{E}.
\end{equation}
The second term on the right hand side of equation \eqref{eq:3.24d} used the fact that when $\dot{r}_{1}=0$, $r_{1}=r_{e}$.

Substituting equations \eqref{eq:3.24b} and \eqref{eq:3.24d} into equation \eqref{eq:3.24a}, and using the dynamical equation \eqref{eq:2.18}, the action $S_{oe}$ then becomes
\begin{eqnarray}\begin{split}
\label{eq:3.25}
S_{oe}=&-2\pi\sigma\int_{\tau_{o}}^{\tau_{e}}r_{1}^{2}d\tau_{E}+\frac{4}{3}\pi\Lambda\int_{\tau_{o}}^{\tau_{e}}(r_{1}^{3}-r_{o}^{3})\frac{\beta_{D2}}{A_{D}}d\tau_{E}
\\&+\frac{4}{3}\pi\Lambda r_{o}^{3}\frac{\beta_{D2}}{A_{D}}\Big|_{\dot{r}_{1}=0}\int_{\tau_{o}}^{\tau_{e}}d\tau_{E}-8\pi\sigma r_{e}^{2}\int_{\tau_{o}}^{\tau_{e}}d\tau_{E}.
\end{split}
\end{eqnarray}
Equation \eqref{eq:2.20} shows that $\beta_{D2}$ is a monotonic function of the variable $r_{1}$. Thus, we do not need to introduce the Heaviside function in equation \eqref{eq:3.25}. This is different from the case of equation \eqref{eq:3.19}.
Combining equations  \eqref{eq:3.25}, \eqref{eq:2.20}, \eqref{eq:2.21} and \eqref{eq:2.22}, the action $S_{oe}$ in equation \eqref{eq:3.25} can be rewritten as
\begin{eqnarray}\begin{split}
\label{eq:3.25a}
S_{oe}=&-\frac{2}{3}\pi\Lambda r_{o}^{3}(\frac{\chi}{\kappa}-\kappa)\int_{r_{o}}^{r_{e}}r_{1}(1-\chi^{2}r_{1}^{2})^{-1}(1-\frac{r_{1}^{2}}{r_{e}^{2}})^{-1/2}dr_{1}
\\&+\frac{2}{3}\pi\Lambda(\frac{\chi}{\kappa}-\kappa)\int_{r_{o}}^{r_{e}}r_{1}^{4}(1-\chi^{2}r_{1}^{2})^{-1}(1-\frac{r_{1}^{2}}{r_{e}^{2}})^{-1/2}dr_{1}\\&
-2\pi\sigma\int_{r_{o}}^{r_{e}}r_{1}^{2}(1-\frac{r_{1}^{2}}{r_{e}^{2}})^{-1/2}dr_{1}\\&+\big\{\frac{2}{3}\pi\Lambda r_{o}^{3}r_{e}(\frac{\chi}{\kappa}-\kappa)(1-\chi^{2}r_{e}^{2})^{-1}-8\pi\sigma r_{e}^{2}\big\}\\&\times\int_{r_{o}}^{r_{e}}(1-\frac{r_{1}^{2}}{r_{e}^{2}})^{-1/2}dr_{1}.
\end{split}
\end{eqnarray}

Using the following integral formulas:
\begin{equation}
\label{eq:A1}%
\int_{a}^{1}x^{2}(1-x^{2})^{-1/2}dx=\frac{1}{2}a(1-a^{2})^{1/2}+\frac{1}{2}arccos(a),
\end{equation}
\begin{eqnarray}\begin{split}
\label{eq:A3}%
\int \frac{x(1-bx^{2})^{-1/2}dx}{(1-ax^{2})}=&- a^{-1/2}(b-a)^{-1/2}\\&\times arctan(\frac{(a-abx^{2})^{1/2}}{(b-a)^{1/2}}),
\end{split}
\end{eqnarray}
\begin{eqnarray}\begin{split}
\label{eq:A6}%
\int \frac{(1-ax^{2})dx}{(1-bx^{2})^{1/2}}=&\frac{1}{2}b^{-3/2}\Big\{ab^{1/2}x(1-bx^{2})^{1/2}\\&-(a-2b)arcsin(b^{1/2}x)\Big\},
\end{split}
\end{eqnarray}
\begin{eqnarray}\begin{split}
\label{eq:A5}%
\int \frac{x^{4}dx}{(1-ax^{2})(1-bx^{2})^{1/2}}=&a^{-2}\int \frac{(1-ax^{2})dx}{(1-bx^{2})^{1/2}}\\&-2a^{-2}\int \frac{dx}{(1-bx^{2})^{1/2}}\\&+a^{-2}\int \frac{dx}{(1-ax^{2})(1-bx^{2})^{1/2}},
\end{split}
\end{eqnarray}
\begin{eqnarray}\begin{split}
\label{eq:A8}%
&\int_{c}^{b} \frac{dx}{(1-ax^{2})(1-b^{-2}x^{2})^{1/2}}\\&=\frac{-b}{2(1-ab^{2})^{1/2}}\big\{arctan\big(\frac{2c(b^{2}-c^{2})^{1/2}(1-ab^{2})^{1/2}}{ac^{2}b^{2}+b^{2}-2c^{2}}\big)-\pi\big\},
\end{split}
\end{eqnarray}
one can finish the integration in equation \eqref{eq:3.25a}. As a result,  equation \eqref{eq:3.27} can be obtained.

\section{The conical singularity }
\label{sec:E}

If we consider only  the contribution of single I-A-I trajectories, it is straightforward to show that the tunneling rate from region $r_{o}<r_{1}<r_{a}$ into region $r_{1}>r_{b}$ (or $r_{1}>r_{e}$) is given by  $e^{-2|S_{ab}|}$ (or $e^{-2|S_{oe}|}$). Thus, the total tunneling rate from region $r_{o}<r_{1}<r_{a}$ into other classically allowed regions is
\begin{equation}
\label{eq:e1}
\Gamma\approx e^{-2|S_{ab}|}+e^{-2|S_{oe}|}.
\end{equation}
Before specifying the forms of the actions  $S_{ab}$ and $S_{oe}$,  we note that equation \eqref{eq:e1} remains  valid regardless of whether conical singularities  are present. This equation can also be derived from equation \eqref{eq:4.29}, which includes  contributions from both  single and multiple I-A-I trajectories. When the dominant contribution comes from single I-A-I trajectories, the contributions from multiple I-A-I trajectories can be neglected, and equation \eqref{eq:4.29} reduces to equation \eqref{eq:e1}.  Equation \eqref{eq:e1} implies that, to study the effect of conical singularities on bubble tunneling, it is sufficient to consider their influence on the actions $S_{ab}$ and $S_{oe}$.

We first consider the effect of conical singularities on the action $S_{ab}$. In this case, domain walls \uppercase\expandafter{\romannumeral1} and \uppercase\expandafter{\romannumeral2} are located on the opposite sides of the throat. A black hole horizon exists at $r=2M$, as shown in figures~\ref{fig:1a} and~\ref{fig:1b}. (In the special case where $M<M_{s}$ and $r_{1}>2M$, figure 7 in ~\cite{SEA} shows that there is no black hole horizon.)  As domain wall \uppercase\expandafter{\romannumeral1} moves, the spacetime structure evolves over time.   When $r_{1}<\chi^{-1}$, there is no cosmological horizon; however, when $r_{1}>\chi^{-1}$,  a cosmological horizon appears at $r=\chi^{-1}$. The conical singularity at a horizon contributes an Euclidean action proportional to the horizon area. Therefore, normally,  conical singularities affect false vacuum bubble tunneling~\cite{31RI,31PR,PG16,OY16,P16,HAT16}.  The subtracted tunneling action is defined as the difference between the Euclidean action of the nucleated spacetime configuration  and that of the background configuration~\cite{SFVF}. For the action $S_{ab}$, the nucleated spacetime configuration corresponds to the state in which domain wall \uppercase\expandafter{\romannumeral1} sweeps through the classically forbidden region $r_{a}<r_{1}<r_{b}$. In contrast, in the background  configuration, domain wall \uppercase\expandafter{\romannumeral1} remains at the critical point $r_{1}=r_{a}$ (or $r_{1}=r_{b}$)~\cite{EAJ}. The critical points $r_{a}$ and $r_{b}$ always satisfy  $2M<r_{a}<r_{b}<\chi^{-1}$~\cite{WDJ}. This indicates that both the nucleated and the background configurations contain the black hole horizon at $r=2M$ (when $M<M_{s}$, neither configuration contains a black hole horizon), but neither includes a cosmological horizon. In other words, both configurations share the same conical singularity. Since this singularity contributes equally to the Euclidean actions of both configurations, it cancels out in the subtraction. Therefore, the conical singularity does not contribute to $S_{ab}$.

Next, we consider the action $S_{oe}$. According to equation \eqref{eq:3.3},  in the background  configuration, domain wall \uppercase\expandafter{\romannumeral1} remains at the critical point $r_{1}=r_{e}$. The nucleated  configuration is defined by domain wall \uppercase\expandafter{\romannumeral1} sweeping through the classically forbidden region $r_{o}<r_{1}<r_{e}$. Since both configurations share the same spacetime for $r>r_{e}$. Only the region $r_{o}<r<r_{e}$  contributes to  $S_{oe}$. Inside domain wall \uppercase\expandafter{\romannumeral1} is Minkowski spacetime. Hence, in the background configuration, there are no conical singularities in the region $r_{o}<r<r_{e}$. Using the definition of $r_{e}$, one can show
\begin{equation}
\label{eq:e2}
r_{e}-\chi^{-1}=-\frac{(\kappa-\chi)^{2}}{\chi(\kappa^{2}+\chi^{2})}<0.
\end{equation}
Equation \eqref{eq:e2} indicates  $r_{e}<\chi^{-1}$, meaning there is no cosmological horizon in the region $r_{o}<r_{1}<r_{e}$ of the nucleated configuration. Thus,  no conical singularities contribute to $S_{oe}$ either.

To sum up, in the case where the dominate contribution comes from single I-A-I trajectories, conical singularities do not affect bubble tunneling up to leading order. Contributions from multiple I-A-I trajectories represent higher-order corrections. Accordingly, the effect of conical singularities on these higher-order trajectories constitutes only a subleading correction to the tunneling rate and can therefore be neglected. In more complex situations where contributions from various multiple I-A-I trajectories may no longer be subleading, the influence of conical singularities on their amplitudes would need to be carefully taken into account. A detailed analysis of such scenarios is beyond the scope of this work and is left for future investigation.


\begin{thebibliography}{99}

\bibitem{EAJ}
E. Farhi, A. H. Guth and J. Guven, Is it possible to create a universe in the laboratory by quantum tunneling?, Nucl. Phys. B {\bf339} (1990) 417.

\bibitem{KMHK}
K. Sato, M. Sasaki, H. Kodama and K. Maeda, Creation of wormholes by first order phase transition of a vacuum in the early universe, Prog. Theor. Phys. {\bf65} (1981) 1443.

\bibitem{HMKK}
H. Kodama, M. Sasaki, K Sato and K. Maeda, Fate of wormholes created by first-order phase transition in the early universe, Prog. Theor. Phys.  {\bf66} (1981) 2052.

\bibitem{K}
K. Lake, Thin spherical shells, Phys. Rev. D {\bf19} (1979) 2847.

\bibitem{SF}
S. R. Coleman and F. De Luccia, Gravitational effects on and of vacuum decay, Phys.  Rev. D {\bf 21} (1980) 3305.

\bibitem{VVI}
V. A. Berezin, V. A. Kuzmin and I. I. Tkachev,  Dynamics of bubbles in general relativity, Phys. Rev. D {\bf36} (1987) 2919.

\bibitem{SEA}
S. K. Blau, E. I. Guendelman and A. H. Guth, Dynamics of false-vacuum bubbles, Phys. Rev. D {\bf35} (1987) 1747.

\bibitem{JC}
J. D. Brown and C. Teitelboim, Neutralization of the Cosmological Constant by Membrane Creation, Nucl. Phys. B {\bf297} (1988) 787.


\bibitem{AME}
A. Aurilia, M. Palmer and E. Spallucci, Evolution of bubbles in a vacuum, Phys. Rev. D {\bf40} (1989) 2511.

\bibitem{WDJ}
W. Fischler, D. Morgan, and J. Polchinski, Quantization of false-vacuum bubbles: a hamiltonian treatment of gravitational tunneling, Phys. Rev. D {\bf42} (1990) 4042.

\bibitem{SFVF}
S. P. De Alwis, F. Muia, V. Pasquarella and F. Quevedo, Quantum transitions between Minkowski and de Sitter spacetimes, Fortschr. Phys. {\bf68} (2020) 2000069.

\bibitem{W}
W. Israel, Singular hypersurfaces and thin shells in general relativity, Nuovo Cimento {\bf44B} (1966) 1.

\bibitem{JS}
J. B. Hartle and S. W. Hawking, Wave function of the universe, Phys. Rev. D {\bf28} (1983) 2960.

\bibitem{A1}
A. Vilenkin, Birth of inflationary universes, Phys. Rev. D {\bf27} (1983) 2848.

\bibitem{A2}
A. Vilenkin, Quantum creation of the universes, Phys. Rev. D {\bf30} (1984) 509.

\bibitem{AD1}
A. D. Linde, Fate of the false vacuum at finite temperature: Theory and applications, Phys. Lett. B {\bf100} (1981) 37.

\bibitem{AD2}
A. D. Linde, Decay of the false vacuum at finite temperature, Nucl. Phys. B {\bf216} (1983) 421.

\bibitem{OJ}
O. Gould and J. Hirvonen, Effective field theory approach to thermal bubble nucleation, Phys. Rev. D  {\bf104} (2021) 096015.

\bibitem{RJ}
R. Li and J. Wang, Vacuum decay and bubble nucleation in the anti-de Sitter black holes, JHEP {\bf09} (2022) 151.

\bibitem{dc}
D. Saito  and C.-M. Yoo, False vacuum decay in rotating BTZ spacetimes, Phys. Rev. D  {\bf104} (2021) 124037.

\bibitem{vf}
V. Pasquarella and F. Quevedo, Vacuum transitions in two-dimensions and their holographic interpretation, JHEP {\bf05} (2023) 192.

\bibitem{jj}
J. R. Espinosaa and J.-F. Fortin, Vacuum decay actions from tunneling potentials for general spacetime dimension,  JCAP {\bf02} (2023) 023.

\bibitem{31RI}
R. Gregory, I.G. Moss and B. Withers, Black holes as bubble nucleation sites, JHEP {\bf03} (2014) 081.

\bibitem{31PR}
P. Burda, R. Gregory and I. Moss, Vacuum metastability with black holes, JHEP {\bf08} (2015) 114.

\bibitem{PG16}
P. Chen, G. Dom$\grave{\mathrm{e}}$nech, M. Sasaki, D. h. Yeom, Thermal activation of thin-shells in anti-de Sitter black hole spacetime, JHEP {\bf1707} (2017) 134.

\bibitem{OY16}
N. Oshita, J. Yokoyama, Creation of an inflationary universe out of a black hole, Phys. Lett. B {\bf785} (2018) 197.

\bibitem{P16}
K. Pasmatsiou, Tunneling between Schwarzschild–de Sitter vacua, Phys. Rev. D {\bf100} (2019)  125013.

\bibitem{HAT16}
H. Firouzjahi, A. Karami, and T. Rostami, Vacuum decay in the presence of a cosmic string, Phys. Rev. D {\bf101} (2020) 104036.


\bibitem{SW}
S. Weinberg, Cosmology, Cambridge University Press, New York (2008).

\bibitem{SD}
S. Dodelson, Modern cosmology, Academic press Inc., U.S. (2003).

\bibitem{SWH}
S. W. Hawking, Wormholes in spacetime, Phys. Rev. D  {\bf37} (1988) 904.

\bibitem{MJ}
M. Lachi$\mathrm{\grave{e}}$ze-Rey and J. P. Luminet, Cosmic topology, Phys. Rep. {\bf254} (1994) 135.

\bibitem{AB}
A. Anderson and B. S. DeWitt, Does the topology of space fluctuate?, Found. Phys. {\bf16} (1986) 91.

\bibitem{JMHAS}
J. Braden, M. C. Johnson, H. V. Peiris, A. Pontzen and S. Weinfurtner, New semiclassical picture of vacuum decay, Phys. Rev. Lett. {\bf 123} (2019) 031601.

\bibitem{KBMMP}
K. L. Ng, B. Opanchuk, M. Thenabadu, M. Reid and P. D. Drummond, Fate of the false vacuum: finite temperature, entropy, and topological phase in quantum simulations of the early universe, PRX Quantum  {\bf2} (2021) 010350.

\bibitem{DAJDS}
D. Jafferis,  A. Zlokapa, J. D. Lykken, D. K. Kolchmeyer, S. I. Davis, N. Lauk, H. Neven and M. Spiropulu, Traversable wormhole dynamics on a quantum processor, Nature {\bf612} (2022) 51.

\bibitem{C1}
C. Sabin, Quantum detection of wormholes, Sci. Rep. {\bf7} (2017) 716.

\bibitem{kb}
K. Skenderis and B. C. van Rees, Holography and wormholes in 2+1 dimensions, Commun. Math. Phys. {\bf301} (2011) 583–626.

\bibitem{KK1221}
K. Krasnov, Holography and Riemann surfaces, Adv. Theor. Math. Phys. {\bf4} (2000) 929.

\bibitem{S1}
S. Weigert, Topological quenching of the tunnel splitting for a particle in a double-well potential on a planar loop, Phys. Rev. A  {\bf50}  (1994) 4572.

\bibitem{L1}
L. S. Schulman, Techniques and applications of path integration, New York: J. Wiley (1981).

\bibitem{SC1}
S. Coleman, Fate of the false vacuum: Semiclassical theory,  Phys. Rev. D  {\bf15} (1977) 2929.

\bibitem{SC2}
C. G. Callan, Jr. and S. R. Coleman, Fate of the false vacuum. II. First quantum corrections, Phys. Rev. D {\bf16} (1977) 1762.


\bibitem{JM}
J. Maldacena, The large N limit of superconformal field theories and supergravity, Adv. Theor. Math. Phys. {\bf2} (1998) 231.


\bibitem{AK1}
A. Kitaev, Hidden correlations in the Hawking radiation and thermal noise, in Talk Given at the Fundamental Physics Prize Symposium (Kavli Institute for Theoretical Physics (KITP), Santa Barbara, 2014).

\bibitem{AK2}
A. Kitaev, A simple model of quantum holography, in Talks at KITP (Kavli Institute for Theoretical Physics (KITP), Santa Barbara, 2015).


\bibitem{SJ}
S. Sachdev and J. Ye, Gapless spin fluid ground state in a random, quantum Heisenberg magnet, Phys. Rev. Lett. {\bf70} (1993) 3339.

\bibitem{SS}
S. Sachdev, Holographic metals and the fractionalized Fermi liquid, Phys. Rev. Lett. {\bf105} (2010) 151602.

\bibitem{JV}
J. Polchinski and V. Rosenhaus, The spectrum in the Sachdev-Ye-Kitaev model, JHEP {\bf1604} (2016) 001.

\bibitem{JD}
J. Maldacena and D. Stanford, Remarks on the Sachdev-Ye-Kitaev model, Phys. Rev. D {\bf94} (2016) 106002.

\bibitem{AS}
A. Kitaev and S. J. Suh, The soft mode in the Sachdev-Ye-Kitaev model and its gravity dual, JHEP {\bf1805} (2018) 183.

\bibitem{PD}
P. Gao and D. L. Jafferis, A traversable wormhole teleportation protocol in the SYK model, JHEP {\bf07} (2021) 097.

\bibitem{JX}
J. Maldacena and X. L. Qi, Eternal traversable wormhole, arXiv:1804.00491 [hep-th].

\bibitem{SEM}
S. Plugge, é. Lantagne-Hurtubise and M. Franz, Revival dynamics in a traversable wormhole, Phys. Rev. Lett. {\bf124} (2020) 221601.

\bibitem{ZZ}
T.-G. Zhou and P. Zhang, Tunneling through an eternal traversable wormhole, Phys. Rev. B {\bf102} (2020) 224305.

\bibitem{MV}
M. Visser, Traversable wormholes from surgically modified Schwarzschild space-times, Nucl. Phys. B {\bf328} (1989) 203.

\bibitem{SWK}
S.-W. Kim, Schwarzschild—de Sitter type wormhole, Phys. Lett. A {\bf166} (1992) 13.

\bibitem{JFS}
J. P. S. Lemos, F. S. N. Lobo and S. Q. de Oliveira, Morris-Thorne wormholes with a cosmological constant, Phys. Rev. D  {\bf68} (2003) 064004.

\bibitem{DCD}
D. C. Dai, D. Minic and D. Stojkovic, New wormhole solution in de Sitter space, Phys. Rev. D  {\bf98} (2018) 124026.

\bibitem{KH1221}
K. Sato, H. Kodama, M. Sasaki and K.-i. Maeda, Multiproduction of universes by first order phase transition of a vacuum, Phys. Lett. {\bf108B} (1982) 103.

\bibitem{KM1221}
H. Kodama, M. Sasaki, K. Sato, and K.-i. Maeda, Fate of wormholes created by first order phase transition in the early universe, Prog. Theor. Phys. {\bf66} (1981) 2052.

\bibitem{JM1225}
J. M. Maldacena and L. Maoz, Wormholes in AdS, JHEP {\bf02} (2004) 053.

\bibitem{PEO1225A}
P. Betzios, E. Kiritsis and O. Papadoulaki, Euclidean wormholes and holography, JHEP {\bf06} (2019) 042.

\bibitem{PEO1225B}
P. Betzios, E. Kiritsis and O. Papadoulaki, Interacting systems and wormholes, JHEP {\bf02} (2022) 126.

\bibitem{MV1225}
M. V. Raamsdonk, Comments on wormholes, ensembles, and cosmology, JHEP {\bf12} (2021) 156.

\bibitem{DJ1225}
D. Marolf and J. E. Santos, AdS Euclidean wormholes, Class. Quantum Grav. {\bf38} (2021) 224002.

\bibitem{MKThorne}
M. S. Morris and K. S. Thorne, Wormholes in spacetime and their use for interstellar travel: A tool for teaching general relativity,  Am. J. Phys. {\bf56} (1988) 395.

\bibitem{FSNL}
F. S. N. Lobo. Wormholes, warp drives and energy conditions, volume 189. Springer (2017).

\bibitem{CLXY}
C.-H. Hao, L.-X. Huang, X. Su and Y.-Q. Wang, Emergence of negative mass in general relativity, Eur. Phys. J. C {\bf84} (2024) 878.


\bibitem{ZCW}
Z. C. Wu, Gravitational effects in bubble collisions, Phys. Rev. D  {\bf28} (1983) 1898.

\bibitem{ST}
S. J. Huber and T. Konstandin, Gravitational wave production by collisions: more bubbles, JCAP {\bf09} (2008) 022.

\bibitem{RTM}
R. Jinno, T. Konstandin and M. Takimoto, Relativistic bubble collisions — a closer look, JCAP {\bf09} (2019) 035.

\bibitem{JJL1}
J. Braden, J. R. Bond and L. Mersini-Houghton, Cosmic bubble and domain wall instabilities I: parametric amplification of linear fluctuations, JCAP {\bf03} (2015) 007.

\bibitem{JJL2}
J. Braden, J. R. Bond and L. Mersini-Houghton, Cosmic bubble and domain wall instabilities II: fracturing of colliding walls, JCAP {\bf08} (2015) 048.

\bibitem{JJL3}
J. R. Bond, J. Braden and L. Mersini-Houghton, Cosmic bubble and domain wall instabilities III: the role of oscillons in three-dimensional bubble collisions, JCAP {\bf09} (2015) 004.

\bibitem{RM}
R. Gobbetti and M. Kleban, Analyzing cosmic bubble collisions, JCAP {\bf05} (2012) 025.

\bibitem{MK}
M. Kleban, Cosmic bubble collisions, Class. Quantum Grav. {\bf28} (2011) 204008.

\bibitem{TJLE}
T. John, J. Giblin, L. Hui, E. A. Lim and I.-S. Yang, How to run through walls: dynamics of bubble and soliton collisions, Phys. Rev. D {\bf82} (2010) 045019.

\bibitem{UHPP}
U. Weiss, H. Grabert, P. H$\mathrm{\ddot{a}}$nggi and P. Riseborough, Incoherent tunneling in a double well, Phys. Rev. B {\bf35} (1987) 9535.

\bibitem{RHB}
R. H. Brandenberger, Quantum field theory methods and inflationary universe models, Rev. Mod. Phys. {\bf57} (1985) 1.

\bibitem{EJW}
E. J. Weinberg, “Classical solutions in quantum field theory : Solitons and Instantons in High Energy Physics,” Cambridge Monographs on Mathematical Physics, September 2012.




\end{thebibliography}
\end{document}